\documentclass[floatfix,prd,showpacs,showkeys,preprintnumbers,nofootinbib,superscriptaddress,11pt]{revtex4-1}
\usepackage[utf8]{inputenc}
\usepackage[sort&compress]{natbib}
\usepackage[normalem]{ulem}
\usepackage{bm}
\usepackage{times}
\usepackage{amssymb,amsbsy,amsmath,amsfonts}
\usepackage{graphicx}
\usepackage{color}
\usepackage{morefloats}
\usepackage{rotating}
\usepackage{slashed}
\usepackage{subfigure}
\usepackage{multirow}
\usepackage{verbatim}
\usepackage{tabularx}
\usepackage{adjustbox}
\usepackage{overpic}
\usepackage[dvipsnames]{xcolor}
\usepackage{nicefrac}
\usepackage{placeins}


\newcommand{\ddsys}{\ensuremath{D D}}
\newcommand{\lclc}{\ensuremath{\Lambda_c\Lambda_c}}
\newcommand{\tev}{\mathrm{TeV}}
\newcommand{\gev}{\mathrm{GeV}}
\newcommand{\mev}{\mathrm{MeV}}
\newcommand{\fm}{\mathrm{fm}}

\usepackage{hyperref}
\hypersetup{
    colorlinks = true,
    urlcolor   = blue,
    linkcolor  = blue,
    citecolor  = blue,
    pdftitle   = {Correlation functions of DD and Lambda-c Lambda-c charmed-hadron pairs},
    pdfauthor  = {Qi-Wei Yuan, Ming-Zhu Liu},
    pdfsubject = {DD and Lambda-c Lambda-c interactions}
}

\begin{document}

\title{\texorpdfstring{Femtoscopic correlation functions between the lightest charmed hadrons}%
{Correlation functions of DD and Lambda-c Lambda-c charmed-hadron pairs}}

\author{Qi-Wei Yuan}
\affiliation{School of Nuclear Science and Technology, Lanzhou University, Lanzhou 730000, China}

\author{Ming-Zhu Liu}
\email[Corresponding author: ]{liumz@lzu.edu.cn}
\affiliation{Frontiers Science Center for Rare Isotopes, Lanzhou University, Lanzhou 730000, China}
\affiliation{School of Nuclear Science and Technology, Lanzhou University, Lanzhou 730000, China}

\date{\today}

\begin{abstract}
Femtoscopic momentum correlation functions provide a unique probe of the
low-energy interactions between short-lived charmed hadrons, which are
inaccessible to conventional scattering experiments. In this work, we
calculate the correlation functions of the lightest charmed-hadron
pairs---$D^{+}D^{+}$, $D^{0}D^{0}$, $D^{+}D^{0}$, and
$\Lambda_{c}^{+}\Lambda_{c}^{+}$---in Pb--Pb collisions at
$\sqrt{s_{\mathrm{NN}}}=5.02$~TeV. Since no established near-threshold
molecular state is known to dominate these channels, their non-perturbative
dynamics is comparatively simple, making them ideal reference systems for
charmed-hadron interaction studies. The relative source functions are
constructed from the hadronization positions in Parton--Hadron--String
Dynamics (PHSD) simulations and exhibit species-dependent, non-Gaussian
shapes that narrow from central to peripheral collisions. The strong
interactions are described by a one-boson-exchange potential and an
effective-field-theory (EFT) contact interaction for $DD$, and by the
contact interaction for $\Lambda_{c}\Lambda_{c}$, with parameters calibrated
to the available lattice QCD scattering lengths and uncertainties propagated
to all observables. Coulomb interactions and quantum-statistical effects are
incorporated consistently in the predicted correlation functions. We find
that the $D^{+}D^{0}$ channel, being free of Coulomb distortion and
identical-particle interference, offers the cleanest access to the repulsive
$DD$ strong interaction, whereas the $\Lambda_{c}^{+}\Lambda_{c}^{+}$
correlation is most pronounced in peripheral collisions, where the compact
source enhances the low-momentum signal. We note that the $\Lambda_{c}\Lambda_{c}$ interaction strength is calibrated to lattice QCD results obtained at a nonphysical pion mass, and the possible pion-mass dependence of this input is not quantified here.   Within the stated lattice constraints and the PHSD source model, these results provide a benchmark estimate and an analysis framework for forthcoming femtoscopic measurements by the ALICE experiment; the quoted uncertainty bands propagate only the lattice scattering-length input.

\end{abstract}

\maketitle
\raggedbottom
\brokenpenalty=10000

\section{Introduction}

Recent studies of exotic states indicate that non-perturbative QCD effects, such as hadron–hadron interactions~\cite{Chen:2016qju,Oset:2016lyh,Guo:2017jvc,Meng:2022ozq,Liu:2024uxn} and three-body interactions~\cite{Pan:2025vyg}, play an important role in understanding the properties and internal structure of exotic states~\cite{Chen:2016qju,Lebed:2016hpi,Oset:2016lyh,Esposito:2016noz,Dong:2017gaw,Guo:2017jvc,Ali:2017jda,Karliner:2017qhf,Guo:2019twa,Liu:2024uxn,Wang:2025sic,Doring:2025sgb,Dai:2026fkg,Bai:2026atm}. Moreover, hadron--hadron interactions are also indispensable for describing
the ``$\rho$--$\pi$ puzzle'' in the strong decays of charmonium states~\cite{Li:1996yn,Zhao:2010mm,Wang:2012mf},
as well as the branching fractions and CP violation in the weak decays
of heavy hadrons~\cite{Jia:2024pyb,Duan:2024zjv,Yuan:2025pnt}. A proper description of such low-energy hadron--hadron interactions is
therefore essential for bridging QCD and the spectrum, structure, and
decay phenomena of hadronic matter. 

Currently, lattice QCD provides the only model-independent, first-principles
approach to studying hadron--hadron interactions directly from the QCD
Lagrangian~\cite{BMW:2008jgk,Briceno:2017max}. The seminal proposal by L\"uscher relates the discrete energy levels of
two-hadron systems in a finite box to the infinite-volume scattering
phase shifts through a set of model-independent equations~\cite{Luscher:1990ux,Luscher:1991cf}. Alternatively, the HAL QCD method extracts the non-local or local
potential from the Bethe--Salpeter wave functions measured on the
lattice, from which hadron--hadron interactions can
be determined~\cite{Ishii:2006ec,Aoki:2012tk}. Complementary to lattice QCD, hadron--hadron interactions can also be
studied within effective field theories (EFTs)~\cite{Guo:2013sya,Liu:2019tjn} and phenomenological
approaches, such as the one-boson-exchange~(OBE) model~\cite{Tornqvist:1993ng,Chen:2015loa,Liu:2019stu}.
However, in contrast to the \emph{ab initio} nature of lattice QCD,
these approaches rely heavily on experimental input or model-dependent
assumptions about the underlying dynamics.

%low-energy hadron--hadron interactions is necessary for connecting QCD with the spectrum and structure of hadronic matter.  These interactions govern hadronic bound and resonant states and shape the final-state dynamics of near-threshold production reactions.  Heavy-flavor systems combine heavy-quark symmetry with nonperturbative light-quark and gluon dynamics and therefore provide a complementary test of hadronic interaction models.

Conventional scattering experiments are a standard means of generating
the data needed to constrain the unknown parameters of hadron--hadron
interactions.
A typical example is the nucleon force, which has been
determined with high precision from abundant proton--proton and
proton--neutron scattering data~\cite{Epelbaum:2008ga,Epelbaum:2014sza,Lu:2021gsb,Lu:2025syk}.
This approach, however, is applicable only to sufficiently long-lived
particles that can be prepared as beams or targets, and it becomes
impractical for unstable hadrons. For hyperon--nucleon systems, for instance, the available scattering
data remain scarce~\cite{Bai:2026syt}.
Open-charm hadrons are an even more challenging case: their short
lifetimes generally preclude conventional fixed-target measurements
and the production of sufficiently intense secondary beams, rendering
direct two-body scattering measurements impractical.

Intensity-frontier femtoscopic measurements in
high-energy heavy-ion and proton--proton collisions offer a
complementary experimental pathway to low-energy hadron--hadron
interactions~\cite{ALICE:2020mfd}.
Two-particle femtoscopy uses hadron pairs produced in the collisions
and relates their low-relative-momentum correlation function to the
space--time distribution of pair emission and the two-body relative
wave function~\cite{Lisa:2005dd,Fabbietti:2020bfg}.
Under the assumptions of chaotic emission and negligible final-state
Coulomb distortion, the correlation
function is directly sensitive to the $S$-wave scattering length and
effective range through the Koonin--Pratt(KP) or Lednick\'y--Lyuboshitz
formalism~\cite{Koonin:1977fh,Lednicky:1981su}.
The method therefore gives access to final-state interactions
involving short-lived particles, without requiring a beam or target
made from the hadrons of interest, and has been successfully applied
to a variety of systems, from pion--pion and kaon--proton pairs to
hyperon--nucleon and hyperon--hyperon pairs~\cite{ALICE:2018ysd,ALICE:2019eol,ALICE:2019hdt,ALICE:2019buq,ALICE:2019gcn}, and even to pairs involving charmed hadrons~\cite{ALICE:2022enj,ALICE:2024bhk}. It should be noted that off-shell ambiguities in the
strong interaction are mainly associated with the short-distance behavior of
the relative wave function and may therefore induce theoretical
uncertainties in correlation function calculations~\cite{Epelbaum:2025aan}. For the near-threshold channels,
however, the phenomenological impact of these ambiguities is expected to be
limited, provided that the low-energy scattering input is fixed. Quantitative two-body studies within realistic interaction models for the systems of interest
support this expectation, finding relatively mild off-shell effects
\cite{Gobel:2025afq,Molina:2025lzw}.

The rapid development of femtoscopic measurements at the Large Hadron Collider~(LHC), in
particular those performed by the ALICE experiment, has demonstrated
the strong potential of this technique for probing interactions
involving charmed hadrons~\cite{ALICE:2022enj}.
This has motivated a series of theoretical studies of systems such as charmed-meson--light-meson, 
hidden charmed hadron pairs,  and doubly charmed hadron
pairs~\cite{Haidenbauer:2020kwo,Torres-Rincon:2023qll,Kamiya:2022thy,Vidana:2023olz,Liu:2023uly,Liu:2023wfo,Shen:2025qpj},
whose interactions are closely connected to the nature of charmed
exotic states.
It should be noted that the underlying non-perturbative dynamics is
further complicated by coupled-channel effects and bare-state dressing
effects~\cite{Shen:2025qpj}, which introduce considerable uncertainties into the
theoretical predictions.  In this work, we focus on systems formed by pairs of the lightest charmed hadrons, i.e., \ddsys\ and \lclc\ \footnote{The $D\Lambda_c$ system is not investigated here,
as it is expected to couple  to excited doubly charmed
baryons, which would substantially complicate the non-perturbative
dynamics of the system.}, which 
 offer the following
distinct advantages:
\begin{enumerate}
\item both systems are composed of the lightest charmed mesons and
    baryons, respectively, and are therefore readily accessible in
    current experimental measurements~\cite{ALICE:2018lyv,ALICE:2018hbc};
    \item their threshold regions are comparatively clean, as no
    established near-threshold molecular state is known to dominate
    the \ddsys\ or \lclc\ threshold~\cite{Li:2012ss,Liu:2019stu,Liu:2026yhn};
    \item lattice QCD calculations of the \ddsys\ and \lclc\
    interactions are available, providing reliable constraints on
    their wave functions~\cite{Shi:2025ogt,Xing:2025uai}.
\end{enumerate}  

A further motivation for choosing Pb--Pb collisions is the charm
multiplicity. A hard process producing a single $c\bar c$ pair in a $pp$
collision supplies only one charm quark, so that a \ddsys\ or \lclc\ pair---each
of which contains two charm quarks---can only arise from events with
additional charm production, e.g., through double-parton scattering or
higher-order processes~\cite{LHCb:2012aiv}. Such events are rare, and the
resulting pair yields in $pp$ are correspondingly limited. In Pb--Pb
collisions, the many nucleon--nucleon encounters increase the charm
multiplicity and provide abundant pairs of two charm-carrying
hadrons~\cite{Song:2018xca}. Moreover, the large number of charm pairs per event dilutes the
initial-state correlations between the two charm quarks that would
otherwise distort the correlation function in $pp$
collisions~\cite{Zhao:2026bhh}. 

Motivated by the above considerations, in this work we study the
\ddsys\ and \lclc\ interactions and calculate the corresponding
femtoscopic correlation functions within a single framework.
For the interaction input, we use an OBE potential and an EFT contact
interaction for \ddsys, and the EFT contact interaction for \lclc.
Their parameters are adjusted to reproduce the lattice QCD scattering
lengths (for \lclc, at the available nonphysical pion mass), with
effective ranges and phase shifts providing further tests.
The resulting scattering wave functions are then combined with
source functions that depend on the particle species and are not
assumed to be Gaussian; these sources are taken from the hadronization
positions in Parton--Hadron--String
Dynamics (PHSD) simulations of Pb--Pb collisions at
$\sqrt{s_{\rm NN}} = 5.02~\tev$~\cite{Cassing:2008sv,Bratkovskaya:2011wp}.
With this setup, we calculate the correlation functions of the
$D^+D^+$, $D^0D^0$, $D^+D^0$, and $\Lambda_c^+\Lambda_c^+$ pairs,
including the strong, Coulomb, and quantum-statistical effects.
Comparing these channels helps to separate the strong, Coulomb, and
quantum-statistical contributions, and the correlation strength at
finite relative momentum reflects the finite-range structure of the
interaction beyond the scattering length.
The resulting correlation functions provide benchmark estimates that
can be compared with future femtoscopic measurements of charmed-hadron
pairs at the LHC.

This work is organized as follows. Section~\ref{sec:framework} introduces
the momentum-correlation observable and its two ingredients: the relative
source function and the two-body scattering wave function. We first
introduce a Gaussian reference source and then specify the
species-dependent, non-Gaussian sources obtained from PHSD simulations,
before constructing the wave functions with the appropriate exchange
symmetry and Coulomb boundary conditions. The strong-interaction
inputs---the OBE potential and the EFT contact interaction---are
presented last.  Section~\ref{sec:results} begins with
the calibration of the interactions and the comparison of the resulting
scattering observables with lattice data, followed by an analysis of the
PHSD source distributions and the correlation functions of the
\(D^+D^+\), \(D^0D^0\), \(D^+D^0\), and \(\Lambda_c^+\Lambda_c^+\) pairs.
Throughout this section, we examine the effects of source size, the
Coulomb interaction, and quantum statistics, as well as the choice of
interaction model at fixed scattering length. A brief summary is given in
Sec.~\ref{sec:summary}.

\section{Theoretical framework}
\label{sec:framework}

The two-particle momentum correlation is described by the KP
formula~\cite{Koonin:1977fh,Pratt:1986cc},
\begin{equation}
    C(\bm{k})
    =
    \int d^3r\, S_R(\bm{r})
    \left|\Psi_{\bm{k}}^{(-)}(\bm{r})\right|^2 ,
    \label{eq:koonin_pratt}
\end{equation}
where $S_R(\bm{r})$ denotes the relative source function, normalized such that
$\int d^3r\, S_R(\bm{r})=1$, and $\Psi_{\bm{k}}^{(-)}(\bm{r})$ is the two-body
wave function satisfying the incoming-wave boundary conditions. Here, $\bm{k}$
and $\bm{r}$ represent the relative momentum and separation of the pair in its
rest frame, respectively. The source function characterizes the spatial
distribution of the pair, whereas the wave
function encodes quantum-statistical effects and the final-state
interactions, both strong and electromagnetic. The two ingredients are
specified in turn below, before the strong-interaction models are
introduced.

\subsection{Source functions}
\label{sec:source_functions}

As a reference parametrization, the source function is commonly taken
to be Gaussian,
\begin{equation}
    S_R(\bm{r})
    =
    \frac{1}{(4\pi R^2)^{3/2}}
    \exp\left(-\frac{r^2}{4R^2}\right),
    \label{eq:gaussian_source}
\end{equation}
where $R$ is the source radius. In practice, however, $R$ is not
universal: it depends on the center-of-mass energy and the size of the
colliding system, and may further vary with the final-state species. To
avoid the model uncertainty associated with a fixed Gaussian
parametrization, we generate the source functions of the \ddsys\ and
\lclc\ systems in Pb--Pb collisions at $\sqrt{s_{\rm NN}}=5.02~\tev$
from PHSD simulations~\cite{Cassing:2008sv,Bratkovskaya:2011wp,Song:2018xca},
constructing them from the hadronization points of the two charmed hadrons
in each pair. The hadronization four-positions and four-momenta are Lorentz
transformed to the pair rest frame, denoted by an asterisk. The earlier-produced hadron is then propagated freely to the later
hadronization time $t_{\rm eq}^{*}=\max(t_1^{*},t_2^{*})$,
using its velocity in the pair rest frame:
\begin{equation}
    \bm{x}_{i,\rm eq}^{*}
    =\bm{x}_i^{*}+\frac{\bm{p}_i^{*}}{E_i^{*}}
    \left(t_{\rm eq}^{*}-t_i^{*}\right),\qquad i=1,2.
    \label{eq:equal_time_positions}
\end{equation}
Eq.~\eqref{eq:equal_time_positions} applies to both hadrons, with the
later one remaining at its original position. Since the free propagation
leaves the four-momenta unchanged, the pair rest frame is unaffected by
this step. The source is constructed from the equal-time separation
$r=|\bm{x}_{1,\rm eq}^{*}-\bm{x}_{2,\rm eq}^{*}|$.
Thus the hadronization-time difference enters both the Lorentz
transformation and the subsequent free propagation. The same prescription
is used for both pair species.  The three-dimensional source for each
species is represented by the radial distribution
\begin{equation}
    P(r)=4\pi r^2 S_R(r),
    \qquad
    \int_0^\infty P(r)\,dr=1.
    \label{eq:radial_source}
\end{equation}

The PHSD histograms have bin widths of $0.1~\fm$ for \ddsys\ and
$0.2~\fm$ for \lclc; the fits use the interval $0\leq r\leq20~\fm$.
For each centrality class, the density in bin $i$ is $n_i/(N\Delta r)$,
where $n_i$ is the bin count and $N$ is the total number of pairs.
For the 0--10\%, 30--50\%, and 60--80\% classes,
$N=41173$, $40965$, and $15063$ for \ddsys, and
$N=1972$, $1321$, and $723$ for \lclc.
The histograms are fitted with the generalized-gamma form
\begin{equation}
    P(r)=
    \frac{\alpha}{\lambda^{\beta+1}
    \Gamma\!\left[(\beta+1)/\alpha\right]}
    r^\beta
    \exp\left[-\left(\frac{r}{\lambda}\right)^\alpha\right],
    \label{eq:phsd_source_fit}
\end{equation}
where $\lambda$ sets the length scale, $\alpha$ controls the tail, and
$\beta$ the small-$r$ behavior; the form is normalized to unity for
$\lambda>0$, $\alpha>0$, and $\beta>-1$, and only these three shape
parameters are fitted. The total pair count
and histogram normalization are unchanged, and the fitted density is
normalized over $0\leq r<\infty$, not over the fit interval.
Beyond $20~\fm$, the generalized-gamma tail is extrapolated rather than
set to zero; this range contributes little to the correlation integrals
because \(P(r)\) is already small there, but it is retained for
completeness.  The fitted distribution is then used directly in the
correlation integrals via $S_R(r)=P(r)/(4\pi r^2)$, and its parameters
are held fixed when propagating the lattice scattering-length
uncertainties. Statistical and fit uncertainties of the source, as well
as the uncertainty associated with the fit range and the equal-time
prescription, are not included in the correlation bands.

For orientation, we characterize each tabulated distribution by the
Gaussian-equivalent radius
\begin{equation}
    R_{\rm src}
    =
    \sqrt{\frac{\langle r^2\rangle_{\rm hist}}{6}},
    \qquad
    \langle r^2\rangle_{\rm hist}
    =
    \frac{\sum_i n_i r_i^2}{\sum_i n_i},
    \label{eq:source_radius}
\end{equation}
where $r_i$ is the bin center and the sums run over all bins of the
original PHSD distribution.

\subsection{Relative wave functions}
\label{sec:wave_functions}

The second ingredient in Eq.~\eqref{eq:koonin_pratt} is the complete relative
scattering wave function of the hadron pair, including the background wave
and its interaction-induced modification. For a neutral pair before exchange
symmetrization, retaining the strong interaction only in the $S$ wave gives
\begin{equation}
    \Psi_{\bm{k}}^{(-)}(\bm{r})
    =e^{i\bm{k}\cdot\bm{r}}+\phi_0^{(-)}(k,r)-j_0(kr),
    \qquad \phi_0^{(-)}(k,r)=\frac{u_0^{(-)}(k,r)}{kr},
    \label{eq:neutral_scattering_wave}
\end{equation}
where $j_0$ is the spherical Bessel function and $u_0^{(-)}$ is the reduced
radial solution normalized to the incoming-wave boundary condition, i.e.,
to a $\delta$-function in momentum.  The difference $\phi_0^{(-)}-j_0$
replaces only the free $S$ wave, while the plane-wave term retains the full
free background. When the interaction vanishes, $\phi_0^{(-)}=j_0$ and the
wave function reduces to a plane wave, yielding $C(k)=1$ for a
distinguishable pair. The radial equation and asymptotic matching are
specified below; the corresponding Coulomb-background replacement and
exchange symmetrization are given in Sec.~\ref{sec:qs_coulomb}.

\subsubsection{Radial wave functions and boundary conditions}
\label{sec:radial_wave_functions}

For each two-body channel, the reduced radial wave function $u_{l}(r)$ is
governed by the Schr\"odinger equation
\begin{equation}
    \left[
        \frac{d^2}{dr^2}
        -\frac{l(l+1)}{r^2}
        +k^2
    \right]u_l(r)
    =
    2\mu V(r)u_l(r),
    \label{eq:schrodinger}
\end{equation}
where $\mu$ is the reduced mass of the pair, $\mu=m_1m_2/(m_1+m_2)$, with
$m_1$ and $m_2$ denoting the masses of the two hadrons. The potential includes
both the strong and the Coulomb contributions,
$V(r)=V_{\rm strong}(r)+V_{\rm Coul}(r)$.  The strong potentials are specified in
Sec.~\ref{sec:strong_interactions}. In the numerical implementation, we use
$\hbar c=197.3~\mev\cdot\fm$ to convert between fm and inverse-MeV units, and
the equation is solved with the Numerov method on a radial grid extending
to $r_{\rm max}=80$~fm with a step size of $0.01$~fm~\cite{Yalikun:2021bfm,Yalikun:2021dpk}. When charged particles
are involved, the Coulomb interaction is included separately as
\begin{equation}
    V_{\rm Coul}(r)=\alpha_{\rm em}\frac{q_1q_2}{r},
    \label{eq:coulomb}
\end{equation}
where $q_i$ is the charge of particle $i$ in units of the elementary charge
and $\alpha_{\rm em}=1/137$ is used consistently in both the potential and
the Coulomb wave functions. 

At large distances, the reduced radial wave function in a neutral channel
is matched to phase-normalized incoming and outgoing free radial waves,
\begin{equation}
    u_l(r)\xrightarrow[r\to\infty]{}
    A_l \widehat{h}_l^{(-)}(kr)+B_l \widehat{h}_l^{(+)}(kr),
    \label{eq:asymptotic}
\end{equation}
where $\widehat{h}_l^{(\pm)}(x)=\pm i x h_l^{(\pm)}(x)$ and
$h_l^{(\pm)}(x)=j_l(x)\pm i y_l(x)$.
For the $S$ wave, this convention gives
$\widehat{h}_0^{(\pm)}(x)=e^{\pm ix}$.
For charged channels, the asymptotic basis is instead given by
the Coulomb radial waves,
\begin{equation}
    H_l^{(\pm)}(\eta,kr)
    =G_l(\eta,kr)\pm iF_l(\eta,kr),
    \label{eq:coulomb_waves}
\end{equation}
where $F_l$ and $G_l$ are the regular and irregular Coulomb functions,
respectively, and
\begin{equation}
    \eta=\frac{\alpha_{\rm em}\mu q_1q_2}{k}
    \label{eq:sommerfeld_parameter}
\end{equation}
is the Sommerfeld parameter. The charged-channel matching reads
\begin{equation}
    u_l(r)\xrightarrow[r\to\infty]{}
    A_l H_l^{(-)}(\eta,kr)+B_l H_l^{(+)}(\eta,kr).
    \label{eq:coulomb_asymptotic}
\end{equation}
The neutral limit satisfies
$H_l^{(\pm)}(0,x)=\widehat{h}_l^{(\pm)}(x)$.
With these conventions, the elastic $S$-matrix element is
\begin{equation}
    S_l=-\frac{B_l}{A_l}=e^{2i\delta_l},
    \label{eq:smatrix}
\end{equation}
and the phase shift is extracted as
\begin{equation}
    \delta_l(k)=\frac{1}{2}\arg S_l(k).
    \label{eq:phase_shift}
\end{equation}
For charged channels, $\delta_l$ denotes the Coulomb-modified strong
phase shift, with the pure Coulomb phase factored out.

\subsubsection{Particle-basis channels and exchange symmetry}
\label{sec:exchange_symmetry}

In the isospin-symmetric limit, the strong interaction between hadrons is
diagonal in the isospin basis, whereas experimentally accessible
observables, such as the momentum correlation function, are defined in
the particle basis.   Relating the two bases is therefore essential for
applying a common \ddsys\ strong interaction to the physical channels. The $D$-meson isodoublet is
\begin{equation}
    D=
    \begin{pmatrix}
        D^+\\
        D^0
    \end{pmatrix},
    \label{eq:d_isodoublet}
\end{equation}
with $I_3(D^+)=+1/2$ and $I_3(D^0)=-1/2$, and the two-body isospin
decomposition
\begin{equation}
    \frac{1}{2}\otimes\frac{1}{2}=1\oplus0.
    \label{eq:isospin_decomposition}
\end{equation}
For the mixed-charge channel, the two particles are interchangeable, and
it is the exchange symmetry, rather than the ordering of the labels,
that carries physical meaning. The basis states with fixed labels,
$|D^+(1)D^0(2)\rangle$ and $|D^0(1)D^+(2)\rangle$, are therefore combined
into the simultaneous eigenstates of the exchange operator $P_{12}$.
Denoting the two-body isospin states by $|I,I_3\rangle$, the connection
between the isospin basis and the particle basis reads
\begin{equation}
\begin{array}{lcl}
    |1,1\rangle  & = & |D^+D^+\rangle,\\[2mm]
    |1,0\rangle  & = &
    \frac{1}{\sqrt{2}}\left(|D^+(1)D^0(2)\rangle
    +|D^0(1)D^+(2)\rangle\right),\\[2mm]
    |1,-1\rangle & = & |D^0D^0\rangle,\\[2mm]
    |0,0\rangle  & = &
    \frac{1}{\sqrt{2}}\left(|D^+(1)D^0(2)\rangle
    -|D^0(1)D^+(2)\rangle\right).
\end{array}
    \label{eq:dd_isospin_states}
\end{equation}

Equation~\eqref{eq:dd_isospin_states} shows that the isovector ($I=1$)
state is symmetric and the isoscalar ($I=0$) state antisymmetric under the
exchange of the two particles,
\begin{equation}
    P_{12}|1,0\rangle=+|1,0\rangle,
    \qquad
    P_{12}|0,0\rangle=-|0,0\rangle.
    \label{eq:isospin_exchange_eigenvalues}
\end{equation}
Within the isospin-symmetric description, the $D$ meson is a spin-zero
boson, so the total wave function of the $D^+D^0$ pair must be symmetric
under the exchange of the two particles; since the spin part is trivially
symmetric, this symmetry is carried entirely by the combined isospin and
spatial parts. A spatial partial wave with
orbital angular momentum $l$ contributes a factor $(-1)^l$ under exchange,
\begin{equation}
    P_{12}\Psi_l=(-1)^l\Psi_l.
    \label{eq:dd_partial_wave_exchange}
\end{equation}
Consequently, Bose symmetry requires even partial waves ($l=0,2,\ldots$)
to combine with the symmetric $I=1$ state, and odd partial waves
($l=1,3,\ldots$) with the antisymmetric $I=0$ state. The \ddsys\ strong
interaction considered in this work is restricted to the $S$ wave
($l=0$); the only allowed $D^+D^0$ strong-interaction channel is
therefore the symmetric $|I=1,I_3=0\rangle$ state, since the $I=0$
component is antisymmetric under exchange and cannot combine with the
symmetric $S$-wave spatial function for a pair of identical bosons. The
three particle channels thus share the same $S$-wave potential:
\begin{equation}
    V_{D^+D^+}^{S\text{-wave}}
    =V_{D^0D^0}^{S\text{-wave}}
    =V_{D^+D^0}^{S\text{-wave}}
    =V_{\ddsys}^{I=1}.
    \label{eq:dd_particle_basis_potential}
\end{equation}

Two remarks are in order concerning Eq.~\eqref{eq:dd_particle_basis_potential}.
First, the equality holds only after $S$-wave projection; no such relation is
assumed for unprojected particle-basis matrix elements. Second, the exchange
symmetry that selects the strong partial wave concerns the interaction, not
the detected particles. Although the $D^+D^0$ strong interaction therefore
proceeds through the symmetric $I=1$ channel, the detected $D^+$ and $D^0$
are distinguishable, and their correlation carries no Bose--Einstein
interference term. The strong potential is nevertheless common to all three
particle channels, which differ only in their Coulomb and
quantum-statistical contributions.

%This assignment is consistent with the
%interpolating-operator construction of Ref.~\cite{Shi:2025ogt}, where the
%symmetric $I=1$ combination is analyzed in the positive-parity even-wave
%sector and the antisymmetric $I=0$ combination in the negative-parity odd-wave
%sector, and it fixes the strong $S$-wave interaction entering each particle-basis correlation function discussed below.

\subsubsection{Quantum statistics and Coulomb-distorted correlations}
\label{sec:qs_coulomb}

The role of quantum statistics  (QS) can be seen most transparently in the absence
of final-state interactions. For identical particles, the exchange symmetry
of the full state---including its spin and other internal components---
determines whether the spatial wave function is symmetric or antisymmetric.  The corresponding free spatial wave functions are
\begin{equation}
    \Psi_{\bm{k}}^{\rm sym/asym}(\bm{r})
    =\frac{1}{\sqrt{2}}
    \left(e^{i\bm{k}\cdot\bm{r}}\pm e^{-i\bm{k}\cdot\bm{r}}\right),
    \label{eq:sym_plane_wave}
\end{equation}
where $\bm{r}$ is the relative coordinate in the pair rest frame. Consequently,
\begin{equation}
    \left|\Psi_{\bm{k}}^{\rm sym/asym}(\bm{r})\right|^2
    =1\pm\cos(2\bm{k}\cdot\bm{r}).
    \label{eq:sym_plane_wave_squared}
\end{equation}
Substituting Eq.~\eqref{eq:sym_plane_wave_squared} into
Eq.~\eqref{eq:koonin_pratt} gives
\begin{equation}
    C_{\rm QS}^{\rm sym/asym}(k)
    =1\pm\int d^3r\,S_R(\bm{r})\cos(2\bm{k}\cdot\bm{r}),
    \label{eq:cqs_fourier}
\end{equation}
which, for the Gaussian source in Eq.~\eqref{eq:gaussian_source}, reduces to
\begin{equation}
    C_{\rm QS}^{\rm sym/asym}(k)=1\pm\exp(-4k^2R^2).
    \label{eq:cqs_gaussian}
\end{equation}

For identical spin-zero $D$ mesons, the spatial state is symmetric,
so $C_{\rm QS}^{\rm sym}(k\to0)=2$, in agreement with the
Bose--Einstein enhancement observed in the measured $K^+K^+$ correlation
functions~\cite{ALICE:2015hvw}. The mixed-charge $D^+D^0$ pair, being
distinguishable (Sec.~\ref{sec:exchange_symmetry}), carries no exchange
interference, and $C(k)=1$ in the absence of final-state interactions. For two identical $\Lambda_c$ baryons, the exchange symmetry of the full
state links the spin and spatial components: the spin-singlet ($S=0$)
state is antisymmetric in spin and pairs with the symmetric $S$-wave
spatial function, whereas the spin-triplet ($S=1$) state is symmetric
in spin and requires an antisymmetric spatial function, which starts
from odd partial waves and thus carries no $S$-wave strong interaction.
In the present approximation, no strong odd-partial-wave interaction is
introduced, and the spin-triplet contribution is treated as a
QS+Coulomb background with an antisymmetric spatial wave function.

For the spin-resolved correlations, the superscripts $\mathrm{Coul}$
and $\mathrm{Coul+strong}$ specify the dynamical interactions retained;
the appropriate spatial symmetrization or antisymmetrization is always
included. By contrast, the auxiliary correlation $C_{\rm Coul}$
introduced below excludes both quantum statistics and the strong interaction. We now turn to the channels with Coulomb repulsion, where the
Coulomb-plus-quantum-statistical background is evaluated separately and
combined with the strong-interaction-modified $S$-wave contribution. The pure
Coulomb part is evaluated with the incoming-wave Coulomb scattering solution
\cite{Mihaylov:2018rva}
\begin{equation}
    \psi_{\bm{k}}^{\rm Coul}(\bm{r})
    =
    e^{-\pi\eta/2}
    \Gamma(1-i\eta)
    e^{i\bm{k}\cdot\bm{r}}\,
    {}_1F_1\!\left(i\eta,1,-i(kr+\bm{k}\cdot\bm{r})\right).
    \label{eq:coulomb_scattering_wave}
\end{equation}
For a symmetric spatial state, the Coulomb wave function is
\begin{equation}
    \psi_{\bm{k}}^{{\rm Coul},{\rm sym}}(\bm{r})
    =
    \frac{
    \psi_{\bm{k}}^{\rm Coul}(\bm{r})+\psi_{\bm{k}}^{\rm Coul}(-\bm{r})
    }{\sqrt{2}}.
    \label{eq:coulomb_sym_wave}
\end{equation}
This symmetric form applies to charged identical $D$-meson pairs and to the
spatial part of the spin-singlet $\Lambda_c^+\Lambda_c^+$ channel. For a
symmetric charged channel, the strong $S$-wave interaction is included by
replacing the pure Coulomb $S$-wave component with the radial solution of the
combined Coulomb and strong potential: for $D^+D^+$, the strong part is the
$I=1$ $DD$ interaction of Sec.~\ref{sec:exchange_symmetry}, and for the
$S=0$ $\Lambda_c^+\Lambda_c^+$ channel it is the corresponding \lclc\
potential specified in Sec.~\ref{sec:strong_interactions}. Defining
\begin{equation}
    \mathcal{A}_{\rm Coul}^{\rm sym}(k,r)
    =
    \frac{1}{2}\int_{-1}^{1}dz\,
    \left|\psi_{\bm{k}}^{\rm Coul,sym}(\bm{r})\right|^2,
    \qquad \bm{k}\cdot\bm{r}=krz,
    \label{eq:coulomb_symmetric_average}
\end{equation}
and using the radial source $P(r)$ from Eq.~\eqref{eq:radial_source} with
$\rho=kr$, the correlation for the symmetric charged channel reads
\begin{equation}
    C_{S=0}^{\rm Coul+strong}(k)
    =
    \int_0^{r_{\rm max}}dr\,P(r)
    \left[
    \mathcal{A}_{\rm Coul}^{\rm sym}(k,r)
    -2\left|\frac{F_0(\eta,\rho)}{\rho}\right|^2
    +2\left|\frac{u_0^{\rm Coul+strong}(k,r)}{\rho}\right|^2
    \right],
    \label{eq:coulomb_strong_swave_replacement}
\end{equation}
where $F_0(\eta,\rho)$ denotes the regular Coulomb $S$ wave and $u_0^{\rm Coul+strong}$
is the radial solution obtained from $V_{\rm Coul}+V_{\rm strong}$. The factors of two
follow from the symmetric identical-particle $S$-wave normalization.
Equation~\eqref{eq:coulomb_strong_swave_replacement} retains all higher
Coulomb partial waves while replacing only the Coulomb $S$-wave component,
thereby avoiding double counting; it should therefore not be interpreted as a
direct sum of independently calculated Coulomb and strong correlation
functions. For the spin-triplet $\Lambda_c^+\Lambda_c^+$ sector, the Coulomb
wave function is instead antisymmetrized,
\begin{equation}
    \psi_{\bm{k}}^{{\rm Coul},{\rm asym}}(\bm{r})
    =
    \frac{
    \psi_{\bm{k}}^{\rm Coul}(\bm{r})-\psi_{\bm{k}}^{\rm Coul}(-\bm{r})
    }{\sqrt{2}},
    \label{eq:coulomb_antisym_wave}
\end{equation}
because the spin-triplet state is symmetric in spin space and the total
fermionic wave function must be antisymmetric. Neglecting the
odd-partial-wave strong interaction, the spin-triplet contribution is
evaluated with this antisymmetrized Coulomb wave function alone,
\begin{equation}
    C_{S=1}^{\rm Coul}(k)
    =
    \int d^3r\,S_R(\bm{r})
    \left|\psi_{\bm{k}}^{\rm Coul,asym}(\bm{r})\right|^2.
    \label{eq:lclc_triplet_coulomb_correlation}
\end{equation}
The no-strong-interaction singlet reference, $C_{S=0}^{\rm Coul}(k)$,
is obtained from Eq.~\eqref{eq:coulomb_strong_swave_replacement}
by setting $V_{\rm strong}=0$ while retaining spatial symmetrization.

For the spin-averaged $\Lambda_c^+\Lambda_c^+$ correlation, the statistical spin weights
are $1/4$ for $S=0$ and $3/4$ for $S=1$, so that, within the present
approximation, the spin-averaged correlation is
\begin{equation}
    C_{\rm avg}(k)
    =
    \frac{1}{4} C_{S=0}^{\rm Coul+strong}(k)
    +
    \frac{3}{4} C_{S=1}^{\rm Coul}(k).
    \label{eq:lclc_spin_averaged_correlation}
\end{equation}

%Neither charged-particle Coulomb form is applied to the neutral $D^0D^0$ or $D^+D^0$ channels.

%To summarize, the different pair channels isolate different combinations of
%the effects described above. Among the \ddsys\ channels, $D^+D^+$ carries
%both Coulomb repulsion and spatial symmetrization, $D^0D^0$ carries spatial symmetrization and the strong final-state interaction, and the distinguishable $D^+D^0$ pair probes the strong interaction alone. The two spin components of \lclc\ probe the symmetric and antisymmetric spatial sectors, with the strong $S$-wave interaction present only in the former. To make this separation explicit for \ddsys, two auxiliary correlation functions are constructed: the $C_{\rm QS}$ calculation retains only the identical-particle quantum-statistical contribution, whereas the $C_{\rm Coul}$ calculation retains only the repulsive Coulomb potential, each with the strong interaction and all other effects switched off.

\subsection{Strong interactions}
\label{sec:strong_interactions}

The strong potential remains as the input of the wave-function construction.
For \ddsys, we consider two independent descriptions, an OBE potential and an
EFT contact potential. For \lclc, only the EFT contact potential is retained
for the results below. The channel assignments follow
Sec.~\ref{sec:exchange_symmetry}.

\subsubsection{ OBE interactions}
\label{sec:obe_interactions}

In the OBE model, the strong interaction is described by the exchange of
light mesons, with the meson-hadron couplings derived from the effective
Lagrangians of Refs.~\cite{Machleidt:1987hj,Swanson:2003tb}. The exchanged
mesons are restricted by isospin, parity, and angular-momentum conservation.
For the \ddsys\ system, this leaves the $\sigma$, $\rho$, and $\omega$ meson exchange, and  the
corresponding effective Lagrangians are given in Ref.~\cite{Liu:2019zvb}. The
\ddsys\ OBE potential reads
\begin{equation}
    V_{\ddsys}(r)=-g_{\sigma D}^2\,Y(\Lambda,m_\sigma,r)
    +C_\rho\, g_{\rho D}^2\,Y(\Lambda,m_\rho,r)
    +\chi\, g_{\omega D}^2\,Y(\Lambda,m_\omega,r).
    \label{eq:vdd}
\end{equation}
The regulated Yukawa function is defined as
\begin{equation}
    Y(\Lambda,m,r)
    =
    \frac{1}{4\pi r}\left(e^{-m r}-e^{-\Lambda r}\right)
    -\frac{\Lambda^2-m^2}{8\pi\Lambda}e^{-\Lambda r},
    \label{eq:yukawa}
\end{equation}
where $m$ is the exchanged-meson mass and $\Lambda$ is the cutoff of the
monopole form factor. In Eq.~\eqref{eq:vdd}, the scalar (vector) exchange
is attractive (repulsive), $C_\rho$ is the isospin weight of the $\rho$
exchange, and $\chi$ fixes the relative sign of the $\omega$ exchange; for
the $I=1$ $S$-wave \ddsys\ channel, $C_\rho=\chi=1$. The value of $\Lambda$
is fixed by the lattice-QCD calibration described in
Sec.~\ref{sec:lattice_constraints}. The
exchanged-meson masses are~\cite{Liu:2019zvb,Lee:2011rka}
\begin{equation}
    m_\sigma=0.600~\gev,\quad
    m_\rho=0.770~\gev,\quad
    m_\omega=0.780~\gev.
    \label{eq:meson_masses}
\end{equation}
Following the coupling convention of Ref.~\cite{Liu:2019zvb}, we use
\begin{equation}
    g_{\sigma D}=3.4,\qquad
    g_{\rho D}=g_{\omega D}=2.6.
    \label{eq:dd_couplings}
\end{equation}

For the \lclc\ system, we also tested an OBE potential containing
$\sigma$- and $\omega$-exchange contributions
\cite{Lee:2011rka,Lu:2017dvm},
\begin{equation}
    V_{\lclc}(r)
    =
    -g_\sigma^2Y(\Lambda,m_\sigma,r)
    +g_\omega^2Y(\Lambda,m_\omega,r),
    \label{eq:vlclc}
\end{equation}
with the same $m_\sigma$ and $m_\omega$ as in Eq.~\eqref{eq:meson_masses} and
the coupling choice of Ref.~\cite{Liu:2019zvb},
\begin{equation}
    g_\sigma=6.8,\qquad g_\omega=5.8.
    \label{eq:lclc_couplings}
\end{equation}

%Within this fixed-coupling, monopole-regulated setup, we cannot simultaneously
%reproduce the lattice effective range and phase shifts by varying the
%cutoff (see Sec.~\ref{sec:lattice_constraints}). We therefore do not present
%\lclc\ OBE results and use the EFT contact interaction instead.

\subsubsection{  EFT contact interactions}
\label{sec:eft_interactions}

We employ the contact interaction of EFT for both systems, providing an
alternative to the OBE description in the \ddsys\ channel. This approach is
widely applied to hadron-hadron
interactions and the study of hadronic molecules
\cite{Guo:2013sya,Liu:2018zzu,Yang:2020nrt,Dong:2021bvy,Pan:2025vyg}. At
leading order, the EFT interaction contains one-pion exchange and contact
terms.  For the \ddsys\ and \lclc\ systems,  one-pion exchange is
forbidden, so that only the contact contribution remains. Upon projection
onto the physical $S$-wave channel (Sec.~\ref{sec:exchange_symmetry}), the
leading nonderivative operator reduces to a channel-dependent low-energy
constant, one for the $I=1$ \ddsys\ channel and one for the $I=0$, $S=0$
\lclc\ channel. These low-energy constants are determined by the lattice QCD
scattering information, as detailed in Sec.~\ref{sec:lattice_constraints}. 
The coordinate-space contact term is
regulated by the normalized Gaussian \cite{Liu:2018zzu}
\begin{equation}
    \delta_{R_c}^{(3)}(\bm r)=\frac{1}{\pi^{3/2}R_c^3}
    \exp\!\left[-\left(\frac{r}{R_c}\right)^2\right],
    \qquad \int d^3r\,\delta_{R_c}^{(3)}(\bm r)=1,
    \label{eq:eft_contact_regulator}
\end{equation}
where $R_c$ is the regulator length. We fix this model parameter separately
for the two systems, taking $R_{c,\ddsys}=1.0~\fm$ and
$R_{c,\lclc}=0.4~\fm$. Absorbing the
regulator normalization into the coefficient $C_{a,X}$, the potential used in
the calculation is
\begin{equation}
    V_{\rm EFT}^{(X)}(r)
    =-C_{a,X}\exp\!\left[-\left(\frac{r}{R_{c,X}}\right)^2\right],
    \qquad X=\ddsys,\lclc.
    \label{eq:contact_potential}
\end{equation}

%For fixed $R_{c,X}$, Eq.~\eqref{eq:contact_potential} is equivalent to
%$V_{\rm EFT}^{(X)}=\widetilde C_X\delta_{R_{c,X}}^{(3)}$ with
%$\widetilde C_X=-\pi^{3/2}R_{c,X}^3C_{a,X}$; $C_{a,X}$ carries units of energy
%and $R_{c,X}$ is the regulator length. 
%With the fitted $C_a<0$, both contact interactions are repulsive. The
%effective ranges and phase shifts follow from this leading-order truncation
%at fixed $R_c$; higher-derivative operators are not considered. The EFT and
%OBE descriptions are evaluated independently throughout.

\FloatBarrier

\section{Results and discussion}\label{sec:results}
 
\subsection{Lattice-QCD scattering constraints}\label{sec:lattice_constraints}

Lattice QCD provides scattering information for both systems, constraining
the strong interaction independently of the correlation observable
\cite{Morita:2019rph,Kamiya:2019uiw,Kamiya:2021hdb}.  The low-energy amplitude
is described by the effective range expansion (ERE),
\begin{equation}
    k\cot\delta_0(k) = \frac{1}{a_0} + \frac{1}{2}r_0 k^2
    +{\cal O}(k^4),
    \label{eq:ere}
\end{equation}
where the lattice quantities are extracted from finite-volume spectra through
L\"uscher's method \cite{Luscher:1986pf,Luscher:1990ux}.  For \ddsys, lattice
QCD finds a weakly repulsive isovector $I(J^P)=1(0^+)$ interaction: the two
ensembles give $a_0=-0.25(5)(1)~\fm$ at $m_\pi\simeq305~\mev$ and
$a_0=-0.25(5)(8)~\fm$ at $m_\pi\simeq207~\mev$, with central effective ranges
$-2.0(13)(2)~\fm$ and $-4.5(30)(12)~\fm$ that vary more visibly but whose
fitted pion-mass slopes are compatible with zero \cite{Shi:2025ogt}. The
physical-pion-mass extrapolation yields $a_0=-0.25(8)(12)~\fm$ and
$r_0=-5.7(45)(17)~\fm$ \cite{Shi:2025ogt}.  The repulsive character is
further supported by a coupled-channel lattice analysis at
$m_\pi\simeq391~\mev$ \cite{PitangaLachini:2025pxr}, which finds
$a_0=-0.191(34)~\fm$, consistent with the values used here, while its
effective range, $r_0=+0.19(27)~\fm$, remains statistically compatible with
the large-uncertainty results of Ref.~\cite{Shi:2025ogt}.  For \lclc, the
first lattice study of the $I(J^P)=0(0^+)$ channel gives
$a_0=-0.21(4)(8)~\fm$ and $r_0=-0.05(13)(25)~\fm$
\cite{Xing:2025uai}. The single-channel treatment of \lclc\ adopted below is supported by the
lattice calculation itself \cite{Xing:2025uai}: no operator mixing between
the $\lclc$ and $\Xi_{cc}N$ ($I=0$) interpolators was observed in the
finite-volume spectrum, and the analyzed energy range lies far below the
$\Sigma_c\Sigma_c$ threshold (more than $300~\mev$ above the $\lclc$
threshold), which is also kinematically excluded from the relative-momentum
region of the correlation functions computed here.  A chiral-EFT analysis of the same data, extrapolated to
the physical pion mass with $m_\pi$-independent couplings, likewise
finds a repulsive interaction \cite{Liu:2026yhn}, supporting the
assumed weak pion-mass dependence.  Since $a_0$ is
considerably better constrained than $r_0$, we fix the leading
near-threshold strength of each interaction to $a_0$, while $r_0$ and the
finite-energy phase shifts serve as independent diagnostics of the energy
dependence.

\begin{table}[ttt]
\centering
\caption{\label{tab:interaction_variations}
Interaction parameters fitted to the central lattice scattering length and
to its lower and upper limits after combining statistical and systematic errors
in quadrature. The intervals propagate this input uncertainty; no additional
cutoff or coupling variation is applied. The EFT regulator length is fixed
separately for each system. Values are rounded for display.}
\begin{adjustbox}{max width=0.96\linewidth,center}
\begin{tabular}{l|lcccc}
\hline\hline
System & Interaction & \(R_c\) [fm] & Parameter & Central value & Interval \\
\hline
\(\ddsys\) & OBE & --- & \(\Lambda\) [GeV] & \(0.5967\) & \([0.5566,0.6440]\) \\
 & EFT & \(1.0\) & \(C_a\) [MeV] & \(-14.652\) & \([-26.813,-5.433]\) \\
\hline
\(\lclc\) & OBE & --- & \(\Lambda\) [GeV] & \(0.6535\) & \([0.6417,0.6642]\) \\
 & EFT & \(0.4\) & \(C_a\) [MeV] & \(-201.259\) & \([-392.338,-89.891]\) \\
\hline\hline
\end{tabular}
\end{adjustbox}
\end{table}

We use the physical-point \ddsys\ scattering length and the nonphysical-point
\lclc\ scattering length as calibration inputs. For each system
$X=\ddsys,\lclc$, the lattice value $a_{0,X}^{\rm lat}$, with statistical
and systematic errors $\sigma_{a,X,\rm stat}$ and $\sigma_{a,X,\rm syst}$
\cite{Shi:2025ogt,Xing:2025uai}, defines the total input uncertainty and
the calibration interval
\begin{equation}
    \sigma_{a,X}=\sqrt{\sigma_{a,X,\rm stat}^2+\sigma_{a,X,\rm syst}^2},
    \qquad a_{0,X}\in[a_{0,X}^{\rm lat}-\sigma_{a,X},\,a_{0,X}^{\rm lat}+\sigma_{a,X}],
    \label{eq:lattice_a0_interval}
\end{equation}
i.e., $\sigma_{a,\ddsys}=0.1442~\fm$ from $-0.25(8)(12)~\fm$ and
$\sigma_{a,\lclc}=0.0894~\fm$ from $-0.21(4)(8)~\fm$ before rounding. A
single unknown parameter per interaction description is then adjusted so
that the model scattering length reproduces the central target
$a_{0,X}^{\rm lat}$ and both limits $a_{0,X}^{\rm lat}\mp\sigma_{a,X}$:
\begin{equation}
    a_{0,X}^{\rm OBE}(\Lambda_X)=a_{0,X}^{\rm target},
    \qquad a_{0,X}^{\rm EFT}(C_{a,X};R_{c,X})=a_{0,X}^{\rm target},
    \label{eq:lattice_parameter_fit}
\end{equation}
where the OBE couplings and meson masses are held fixed and the monopole
cutoff $\Lambda$ is determined, while for EFT the regulator length is
fixed ($R_{c,\ddsys}=1.0~\fm$, $R_{c,\lclc}=0.4~\fm$) and the contact
coupling $C_{a,X}$ is determined. By monotonicity of $a_0$ in each
parameter, the solution is unique for every target, and the three solutions
define the central value and interval of the fitted parameter, listed in
Table~\ref{tab:interaction_variations}. For each observable $\mathcal{O}(x)$, the
propagated envelope is defined by the pointwise extrema over the fitted
interval. Denoting the central fitted parameter by $p_0$, we define
\begin{equation}
    \Delta_-\mathcal{O}(x)=\mathcal{O}_{p_0}(x)-\mathcal{O}_{\min}(x),\qquad
    \Delta_+\mathcal{O}(x)=\mathcal{O}_{\max}(x)-\mathcal{O}_{p_0}(x),
    \label{eq:cutoff_envelope}
\end{equation}
and the shaded bands span $[\mathcal{O}_{\min}(x),\mathcal{O}_{\max}(x)]$. These envelopes
propagate only the lattice scattering-length input uncertainty, not a
complete theoretical error estimate or a simultaneous confidence band;
source-function, coupling, $R_c$, pion-mass-extrapolation, and
omitted-interaction uncertainties are not included. Each fitted interval is
propagated through the ERE parameters, phase shifts, and correlation
functions.

\begin{table}[ttt]
\centering
\caption{\label{tab:obe_lattice_comparison}
ERE parameters from lattice QCD, the OBE interaction, and the EFT contact
interaction.  All entries use the
convention of Eq.~\eqref{eq:ere}; the lattice entries are taken from
Refs.~\cite{Shi:2025ogt,Xing:2025uai}. Model errors propagate the combined
lattice scattering-length errors through the fitted intervals in
Table~\ref{tab:interaction_variations}. The \ddsys\ model
values use the physical reduced mass, while the \lclc\ model values use the
lattice reduced mass corresponding to the quoted nonphysical-pion-mass
result.  Model results are rounded to the precision of the corresponding
lattice entries after the errors in Eq.~\eqref{eq:cutoff_envelope} are
evaluated. The effective range is not fitted. These errors do not include
other sources of theoretical uncertainty.}
\begin{adjustbox}{max width=0.96\linewidth,center}
\begin{tabular}{l|lcc}
\hline \hline
System & Interaction & \(a_0\) [fm] & \(r_0\) [fm] \\
\hline
\(\ddsys\) & Lattice & \(-0.25(8)(12)\) & \(-5.7(45)(17)\) \\
 & OBE & \(-0.25\pm0.14\) & \(-3.1^{+1.4}_{-4.9}\) \\
 & EFT, \(R_c=1.0~\fm\) & \(-0.25\pm0.14\) & \(-2.5^{+1.5}_{-5.5}\) \\
\hline
\(\lclc\) & Lattice & \(-0.21(4)(8)\) & \(-0.05(13)(25)\) \\
 & OBE & \(-0.21\pm0.09\) & \(-3.01^{+1.08}_{-2.00}\) \\
 & EFT, \(R_c=0.4~\fm\) & \(-0.21\pm0.09\) & \(-0.17^{+0.23}_{-0.57}\) \\
\hline \hline
\end{tabular}
\end{adjustbox}
\end{table}

Table~\ref{tab:obe_lattice_comparison} compares the ERE parameters of the
OBE and EFT interactions with the lattice-QCD inputs.  The \ddsys\ values
are evaluated at the physical reduced mass $\mu_{\ddsys}^{\rm phys}=933.625~\mev$
for comparison with the physical-point extrapolation
\cite{ParticleDataGroup:2024cfk,Shi:2025ogt}, while the \lclc\ values use
$\mu_{\lclc}^{\rm lat}=1206.5~\mev$ at $m_\pi\simeq303~\mev$
\cite{Xing:2025uai}. Both descriptions reproduce the small \ddsys\
scattering length and its prescribed interval; their effective ranges are
less negative than the lattice central value but remain compatible with its
large uncertainty. The EFT regulator lengths are fixed separately for the
two systems: $R_{c,\lclc}=0.4~\fm$ is chosen so that the predicted
$r_0=-0.17~\fm$ falls within the statistical uncertainty of the lattice
result, exploiting the monotonic growth of the effective range of a
regulated contact interaction with the regulator length. The regulator
lengths are held fixed in all propagations below, so the EFT bands reflect
only the fitted $C_a$ intervals; the residual $R_c$ dependence is part of
the unquantified theoretical uncertainty. For \lclc, both descriptions
match the small repulsive scattering length, in agreement with
Ref.~\cite{Liu:2026yhn}; the EFT interaction also reproduces the lattice
effective range within its statistical uncertainty, whereas the OBE value
($-3.01~\fm$, interval $[-5.01,-1.93]~\fm$) requires the large lattice
systematic error for compatibility. This OBE--EFT difference in $r_0$ is
the clearest present indication of the sensitivity to the interaction
shape once the threshold strength is fixed.  Finally, because the $\Lambda_c \Lambda_c$ calibration uses a single, nonphysical pion mass ($m_\pi \simeq 303$~MeV), the physical correlation calculation retains the lattice value of $a_0$. A chiral-EFT analysis of the same lattice data, extrapolated to the physical pion mass with $m_\pi$-independent couplings, suggests that the interaction remains repulsive at the physical point~\cite{Liu:2026yhn}, but the size of any residual pion-mass dependence is not quantified here and is treated as an additional theoretical uncertainty. The $\Lambda_c^+\Lambda_c^+$ correlation functions presented below should therefore be read as benchmark estimates based on the nonphysical-point lattice input, rather than as physical-point predictions; a lattice determination at or near the physical pion mass would be required to remove this limitation.

\begin{figure}[ttt]
    \centering
    \subfigure[~\(\ddsys\), \(m_\pi\simeq207~\mev\).]{
        \includegraphics[width=0.48\textwidth]{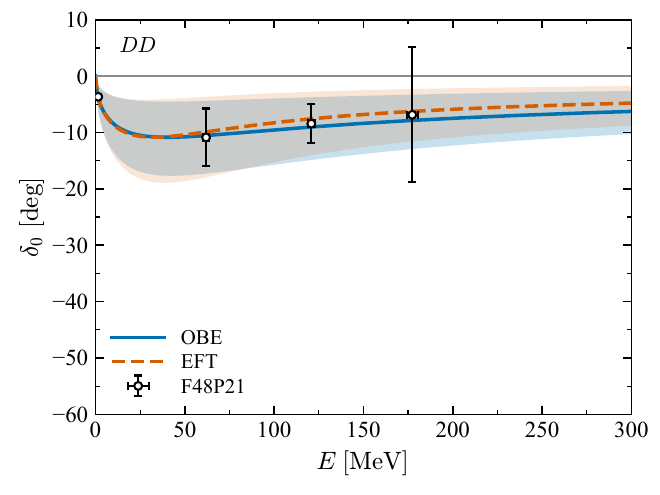}}
    \hfill
    \subfigure[~\(\lclc\), \(m_\pi\simeq303~\mev\).]{
        \includegraphics[width=0.48\textwidth]{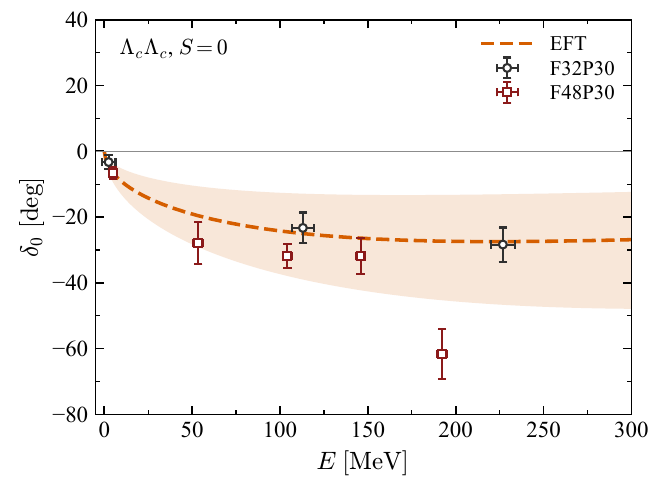}}
    \caption{\label{fig:phase_shift_comparison}
    Strong-interaction \(S\)-wave phase shifts from OBE and EFT for \ddsys\
    in (a), and from EFT for \lclc\ in (b),
    compared with lattice data on the common
    energy axis \(E=k^2/(2\mu)\). The F48P21 points in (a) are digitized and
    adapted from Ref.~\cite{Shi:2025ogt}; the F32P30 and F48P30 points in (b)
    are digitized and adapted from Ref.~\cite{Xing:2025uai}.
    The error bars retain the intervals shown in the original figures,
    with transformed horizontal endpoints. The colored bands propagate
    only the combined lattice scattering-length errors through the model
    parameters. The model reduced masses are \(950.5~\mev\) in (a) and
    \(1206.5~\mev\) in (b); the parameters are not refitted to these
    phase-shift points.}
\end{figure}

Figure~\ref{fig:phase_shift_comparison} overlays the model predictions with
the published lattice phase-shift points on the common axis $E=k^2/(2\mu)$;
the points are digitized from Refs.~\cite{Shi:2025ogt,Xing:2025uai}. For
\ddsys\ at $m_\pi\simeq207~\mev$ [Fig.~\ref{fig:phase_shift_comparison}(a), $\mu=950.5~\mev$], both the
OBE (solid) and EFT (dashed) descriptions are shown: the two central curves
agree near threshold, where the phase shift is governed by the calibrated
scattering length, and remain close to the F48P21 data points over the
measured range, with their mutual separation and band widths growing
gradually toward higher momentum. The individual bands propagate the
$a_0$ input uncertainty, while the separation of the central curves traces
the difference in their predicted effective ranges. For \lclc\ at
$m_\pi\simeq303~\mev$ [Fig.~\ref{fig:phase_shift_comparison}(b), $\mu=1206.5~\mev$], only the EFT result
is shown: with $R_c=0.4~\fm$ its $r_0=-0.17~\fm$ reproduces the mild
energy dependence of the F32P30 and F48P30 points from threshold to
$E\simeq200~\mev$, whereas the OBE interaction is not displayed because,
once calibrated to the \lclc\ scattering length, its
$r_0\simeq-3.0~\fm$ turns the phase shift away from the data. For
comparison, the NLO ChEFT fit of Ref.~\cite{Liu:2026yhn}, which adds
two-pion exchange and is fitted to the phase-shift points directly,
describes the \lclc\ data with a band of comparable width to our envelope,
indicating that different truncations agree at low momentum---where the
calibrated scattering length dominates---and separate at finite momentum,
where the interaction shape matters. Finally, the two \ddsys\ ensembles
give nearly unchanged scattering lengths but visibly different effective
ranges and phase shifts; given the current precision, these are
complementary diagnostics rather than simultaneous fit targets.

\subsection{PHSD source functions}\label{sec:phsd_results}

\begin{figure}[ttt]
    \centering
    \subfigure[~\(\ddsys\).]{
        \includegraphics[width=0.48\textwidth]{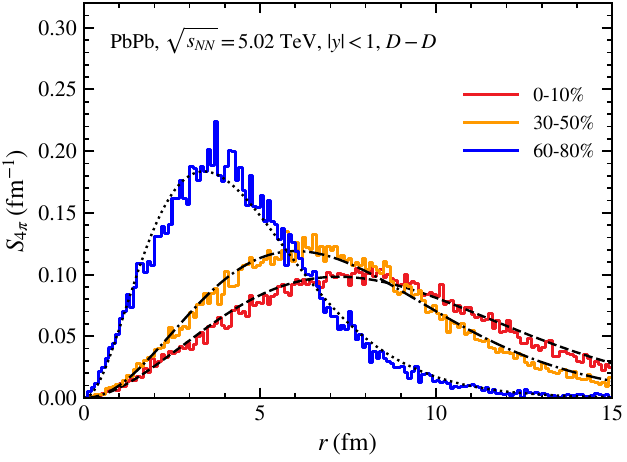}
    }
    \subfigure[~\(\lclc\).]{
        \includegraphics[width=0.48\textwidth]{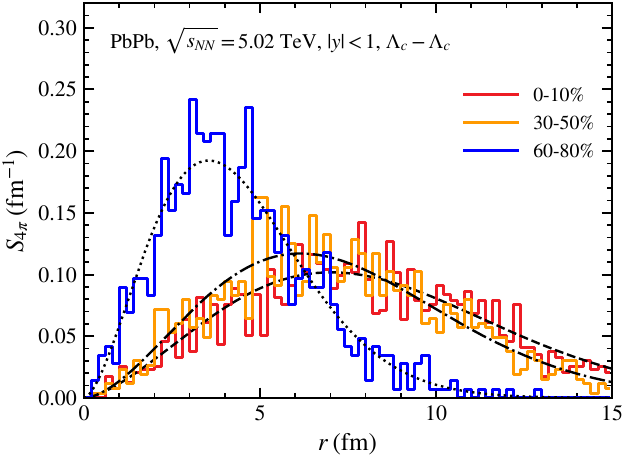}
    }
    \caption{\label{fig:phsd_sources}
    PHSD relative source functions for (a) \ddsys\ and (b) \lclc\ pairs in the
    0--10\%, 30--50\%, and 60--80\% centrality classes.  The plotted radial
    probability density is \(P(r)\equiv S_{4\pi}(r)=4\pi r^2S_R(r)\), with \(r\)
    in fm.  The colored step curves show the PHSD distributions and the black curves
    show generalized-gamma fits.}
\end{figure}

The dedicated PHSD \ddsys\ and \lclc\ source functions and their generalized-gamma
fits are shown in Fig.~\ref{fig:phsd_sources}; the fitted parameters are summarized in
Table~\ref{tab:phsd_source_fit}. In both systems the most central collisions exhibit
the broadest relative-separation distribution, while the source narrows monotonically
toward peripheral events. For \ddsys, the Gaussian-equivalent radii are
\(R_{\rm src}=4.2184\), \(3.4567\), and \(2.1102~\fm\) in the
0--10\%, 30--50\%, and 60--80\% centrality classes, respectively;
the corresponding \lclc\ radii are \(3.9256\), \(3.3499\), and
\(1.9821~\fm\). The two sets of radii differ by about
3--7\% across the three centrality classes.
Because each source is constructed independently from the equal-time
separations of the corresponding pair species, this proximity of scales
is consistent with a common spatial scale inherited from the charm-production
environment and the collision geometry of the PHSD events
\cite{Cassing:2008sv,Bratkovskaya:2011wp}. It should be emphasized, however, that
similar radii do not imply coincident hadronization points: the independently
extracted \lclc\ source remains slightly more compact in every centrality class, and
both distributions retain species-dependent information on production and
hadronization dynamics. Finally, these relative-source scales are smaller than the
single-particle homogeneity volumes of light-flavor pairs obtained in femtoscopic
analyses of Pb--Pb collisions at the same energy \cite{ALICE:2026bgl}. This difference may be associated with the larger transverse masses
of charmed-hadron pairs. In a collectively expanding system,
space--momentum correlations can lead to smaller homogeneity lengths
at larger transverse masses, as observed in light-hadron femtoscopy
\cite{ALICE:2015hvw}. 

\begin{table}[ttt]
\centering
\caption{\label{tab:phsd_source_fit}
Generalized-gamma fit parameters for the PHSD \ddsys\ and \lclc\ equal-time
radial source distributions, fitted over \(0\leq r\leq20~\fm\) with
full-space normalization fixed to unity.
The fitted quantity is \(P(r)=4\pi r^2S_R(r)\), with \(r\) in fm.
The last column gives the Gaussian-equivalent radius \(R_{\rm src}\)
calculated from the complete tabulated distribution using
Eq.~\eqref{eq:source_radius}.}
\begin{adjustbox}{max width=0.96\linewidth,center}
\begin{tabular}{l|lcccc}
\hline\hline
System & Centrality & \(\lambda\) [fm] & \(\alpha\) & \(\beta\) &
\(R_{\rm src}\) [fm] \\
\hline
\multirow{3}{*}{\(\ddsys\)}
 & 0--10\%  & 6.9977 & 1.6793 & 1.7688 & 4.2184 \\
 & 30--50\% & 4.7277 & 1.4794 & 2.1113 & 3.4567 \\
 & 60--80\% & 2.3153 & 1.2405 & 2.0006 & 2.1102 \\
\hline
\multirow{3}{*}{\(\lclc\)}
 & 0--10\%  & 7.9642 & 1.9279 & 1.5101 & 3.9256 \\
 & 30--50\% & 6.3980 & 1.7996 & 1.6758 & 3.3499 \\
 & 60--80\% & 3.6565 & 1.6955 & 1.6058 & 1.9821 \\
\hline\hline
\end{tabular}
\end{adjustbox}
\end{table}

Beyond this common scale, the fitted source shapes exhibit a nontrivial
centrality dependence. For all six fits, \(1<\alpha<2\), so the fitted
generalized-gamma tails fall asymptotically more slowly than a Gaussian
but faster than an exponential. This is a property of the fit extrapolation,
not a quantitative constraint on the PHSD tail beyond the fit interval.
The centrality-dependent peak and small-distance shapes cannot be captured
by a single Gaussian radius. Earlier theoretical studies of charmed-hadron correlation functions
commonly adopted a fixed Gaussian source with a radius of about \(0.8\)--\(1~\fm\),
independent of the collision system and final state
\cite{Haidenbauer:2020kwo,Torres-Rincon:2023qll,Kamiya:2022thy}. The present
species-dependent, non-Gaussian sources eliminate this source of model freedom and,
as quantified in the following sections, the smaller and more compact peripheral
sources strengthen the low-momentum correlation signal. These features motivate
evaluating the KP integral with the full PHSD radial distribution, rather
than with \(R_{\rm src}\) alone or with the Gaussian reference of
Eq.~\eqref{eq:gaussian_source}
\cite{ALICE:2023sjd,Epelbaum:2025aan,Xiong:2025bmd}.

\subsection{\ddsys\ momentum correlation functions}

Once the scattering interactions are fixed, the same parameters are retained for the
physical correlation calculation, while the Schr\"odinger equation is solved with the
physical reduced masses given in Sec.~\ref{sec:radial_wave_functions}. The OBE
interaction or the EFT contact interaction is therefore evaluated with masses
appropriate to physical charmed hadrons, and no refitting is performed when the
resulting wave functions are folded with the PHSD sources. The fitted intervals in
Table~\ref{tab:interaction_variations} are propagated to the correlation functions
without changing the source functions.

\begin{figure}[!htbp]
    \centering
    \subfigure[~Quantum-statistical contribution \(C_{\rm QS}\).]{
        \includegraphics[width=0.46\textwidth]{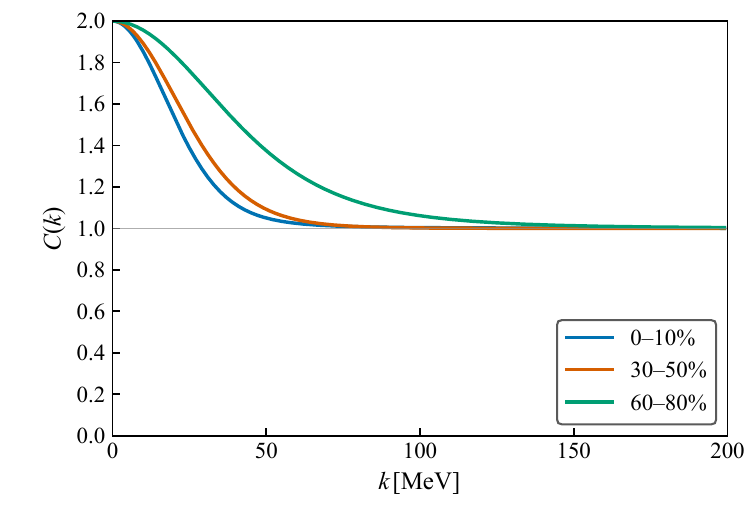}
    }
    \subfigure[~Coulomb contribution \(C_{\rm Coul}\).]{
        \includegraphics[width=0.46\textwidth]{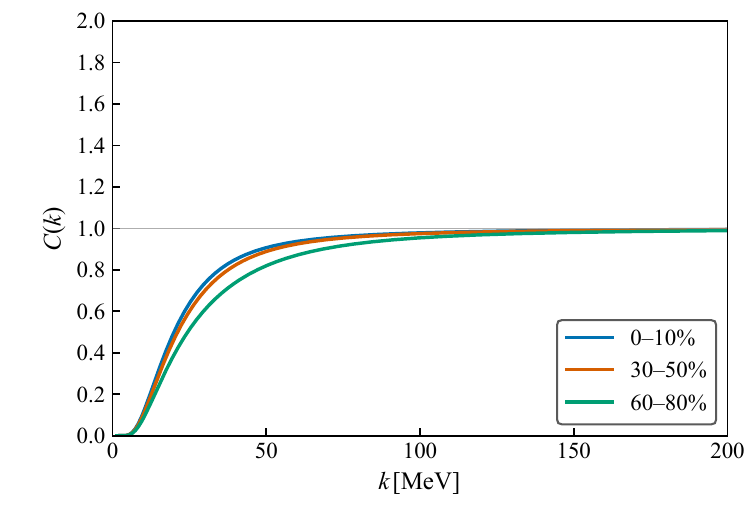}
    }
    \caption{\label{fig:dd_auxiliary_correlations}
    Auxiliary \ddsys\ correlations for the same three centrality classes, obtained
    with the strong interaction switched off. These panels therefore carry no
    propagated scattering-length error band.}
\end{figure}

It is instructive to first inspect the two background ingredients separately, with
the strong interaction switched off, as shown in
Fig.~\ref{fig:dd_auxiliary_correlations}. As shown in Fig.~\ref{fig:dd_auxiliary_correlations}(a), quantum
statistics alone gives \(C_{\rm QS}^{\rm sym}(0)=2\) for an
exchange-symmetric spatial state and any normalized source, while
its finite-momentum dependence reflects the source distribution.
By contrast, the Coulomb-only reference in
Fig.~\ref{fig:dd_auxiliary_correlations}(b), with quantum statistics
and the strong interaction switched off, is suppressed to zero
as \(k\to0\) by Coulomb repulsion and approaches unity from below
as the relative momentum increases. The \(C_{\rm
Coul}\) panel shows the opposite low-momentum behavior generated by Coulomb
repulsion: a strong suppression at small \(k\) that relaxes to unity from below as
the Sommerfeld distortion weakens. These two panels define the model-independent
backgrounds against which the full correlation functions are to be read.

\begin{figure}[!htbp]
    \centering
    \subfigure[]{
        \includegraphics[width=0.46\textwidth]{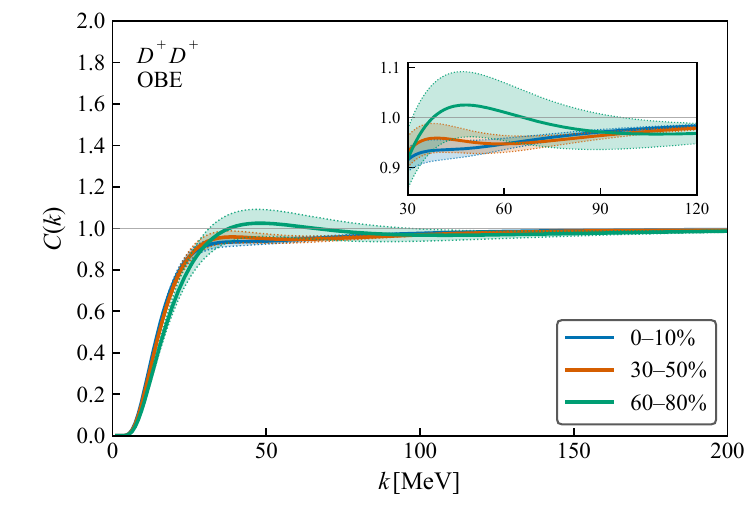}
    }
    \subfigure[]{
        \includegraphics[width=0.46\textwidth]{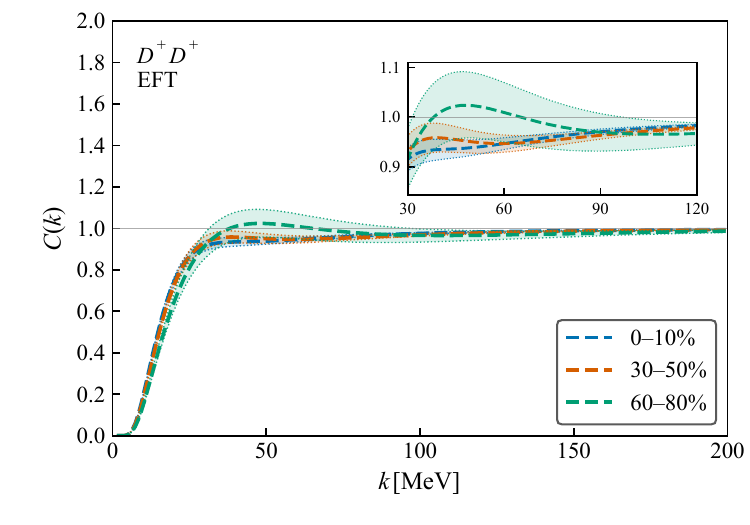}
    }
    \\[-0.2em]
    \subfigure[]{
        \includegraphics[width=0.46\textwidth]{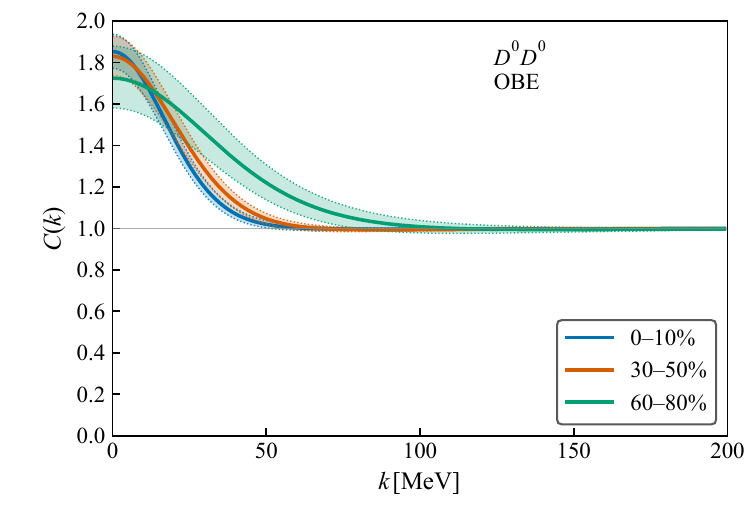}
    }
    \subfigure[]{
        \includegraphics[width=0.46\textwidth]{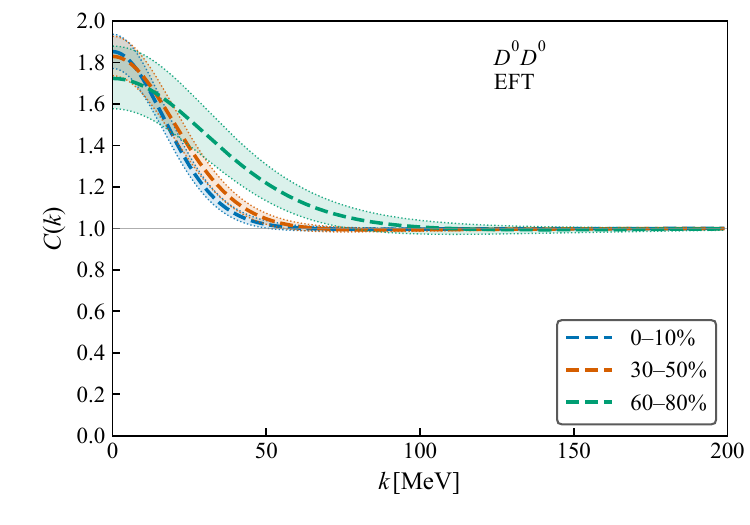}
    }
    \\[-0.2em]
    \subfigure[]{
        \includegraphics[width=0.46\textwidth]{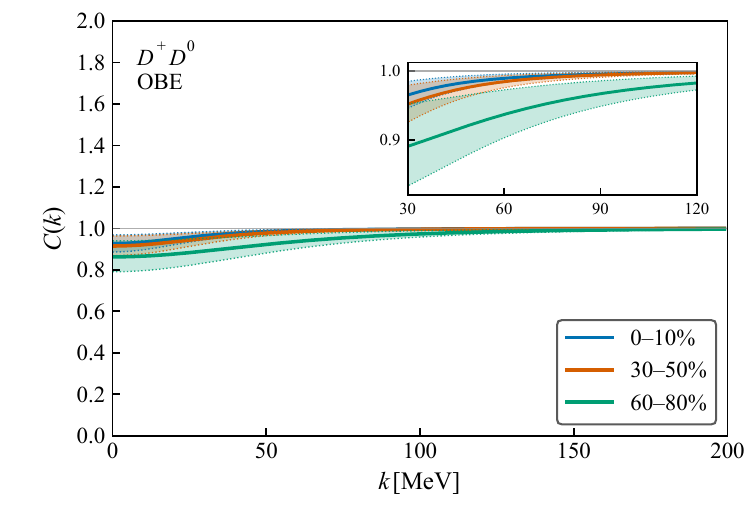}
    }
    \subfigure[]{
        \includegraphics[width=0.46\textwidth]{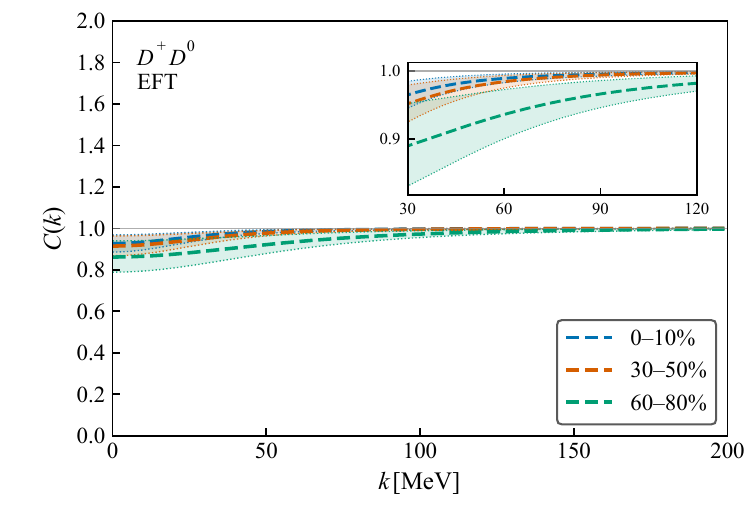}
    }
    \vspace{-0.4em}
    \caption{\label{fig:dd_correlation_models}
    Physical \ddsys\ momentum correlation functions obtained with the PHSD source
    functions for 0--10\%, 30--50\%, and 60--80\% centralities.
    Rows show \(D^+D^+\), \(D^0D^0\), and \(D^+D^0\), while columns show OBE
    and EFT.
    The central parameters and fitted intervals are given in
    Table~\ref{tab:interaction_variations}. Shaded bands propagate the combined
    lattice scattering-length errors, with the source functions held fixed.
    OBE central curves are solid and EFT central curves are dashed; colors
    distinguish the centralities. The insets magnify \(C(k)\) in the
    \(30\)--\(120~\mev\) region.}
\end{figure}

Figure~\ref{fig:dd_correlation_models} shows the full \ddsys\ momentum correlations
from the OBE interaction and from the EFT contact interaction with \(R_c=1.0~\fm\).
The three channels respond differently to the strong interaction because they carry
different combinations of the two backgrounds established above. For the identical charged channel \(D^+D^+\), shown in Figs.~\ref{fig:dd_correlation_models}(a) and \ref{fig:dd_correlation_models}(b), \(C(k)\) is shaped by
competition between Coulomb suppression at very small momentum and the
quantum-statistical enhancement at finite momentum: the correlation is
suppressed below unity where the Coulomb distortion of
Fig.~\ref{fig:dd_auxiliary_correlations}(b) dominates, and the
Bose--Einstein term lifts it above unity once the Coulomb effect weakens,
so that \(C(k)\) approaches unity from above. The repulsive strong
final-state interaction enters only as a subleading modification of this
structure: it adds a slight further suppression at very low \(k\), where
the Coulomb distortion dominates, and reduces the height of the
Bose--Einstein peak at intermediate \(k\).

The \(D^0D^0\) channel, shown in Figs.~\ref{fig:dd_correlation_models}(c) and \ref{fig:dd_correlation_models}(d), carries no Coulomb interaction, so its low-\(k\) structure
originates entirely from the quantum-statistical background of
Fig.~\ref{fig:dd_auxiliary_correlations}(a), modified by the repulsive \ddsys\
strong interaction. Quantum statistics alone would give \(C_{\rm QS}(0)=2\), but
the repulsive final-state interaction depletes the overlap between the symmetrized
wave function and the source at short distance; consequently, the low-momentum correlation lies below 2, with the suppression becoming stronger toward peripheral collisions. As \(k\) increases,
the correlation decreases, passes slightly below unity, and then
returns toward unity. The height of the low-\(k\) peak thus encodes the
interplay between Bose--Einstein enhancement and repulsive strong scattering, with
the turnover position set predominantly by the source radius.

The \(D^+D^0\) pair, shown in Figs.~\ref{fig:dd_correlation_models}(e) and \ref{fig:dd_correlation_models}(f), carries neither background: the detected \(D^+\) and
\(D^0\) are distinguishable, so no identical-particle interference term
appears, and the \(D^0\) is neutral, so the two-body Coulomb interaction
is absent. The exchange symmetry of the isospin wave function at \(l=0\)
still selects the symmetric \(I=1\) two-\(D\) state, and the \(D^+D^0\)
final-state interaction is governed by the same \(I=1\) \ddsys\
\(S\)-wave potential as the identical-charge channels. The correlation is therefore
driven solely by the strong interaction: the calibrated repulsive interaction
suppresses \(C(k)\) below unity over the finite-\(k\) region, and \(C(k)\) rises
monotonically toward unity from below. The deviation is small, of order a few
percent, simply because the calibrated scattering length is small; the value of
this channel lies precisely in the absence of any competing background. Among the
three channels, \(D^+D^+\) contains all three ingredients, \(D^0D^0\) isolates the
interplay of quantum statistics and the strong interaction, and \(D^+D^0\) probes
the strong interaction alone.

The centrality dependence of all three channels follows directly from
the source distributions of Sec.~\ref{sec:phsd_results}: the compact
60--80\% source weights the short relative separations at which the
interaction-modified wave function differs most from its free
counterpart, thereby enhancing the entire low-\(k\) structure. The
effect is largest in absolute terms for the \(D^+D^+\) and \(D^0D^0\)
channels, where the signal itself is sizeable. In the low-momentum region, the OBE and EFT benchmark estimates remain
close, since both are anchored to the same calibrated scattering length;
the shaded bands, which propagate the lattice uncertainty of \(a_0\),
would be the dominant quantified uncertainty in a comparison with data,
although source and interaction-shape uncertainties are not included. At finite relative momentum, the small residual differences between
the central curves reflect sensitivity to interaction information
beyond the scattering length, including their different effective
ranges (Table~\ref{tab:obe_lattice_comparison}). Sufficiently precise measurements of the finite-momentum dependence, together with controlled source uncertainties, could provide constraints on interaction information beyond the scattering length, including the effective range.

\subsection{\texorpdfstring{\lclc\ momentum correlation functions}{Lambda-c Lambda-c momentum correlation functions}}

For two identical $\Lambda_c^+$ baryons, fermionic antisymmetry ties the spin
state to the exchange symmetry of the spatial wave function. The spin-singlet
$S=0$ state is antisymmetric in spin and must therefore combine with a
symmetric spatial wave function; in the present $S$-wave approximation this is
the only component that carries the calibrated $\Lambda_c\Lambda_c$ strong
interaction. The spin-triplet $S=1$ state is symmetric in spin and requires
an antisymmetric spatial wave function, whose lowest allowed partial wave is
the \(P\) wave. Since no odd-partial-wave strong interaction is introduced
here, the $S=1$ sector is treated as a background channel and is evaluated
with the antisymmetrized Coulomb wave function alone. The strong interaction
thus enters the spin-averaged observable only through the $S=0$ component,
on top of the repulsive Coulomb background.

\begin{figure}[ttt]
    \centering
    \subfigure[]{
        \includegraphics[width=0.46\textwidth]{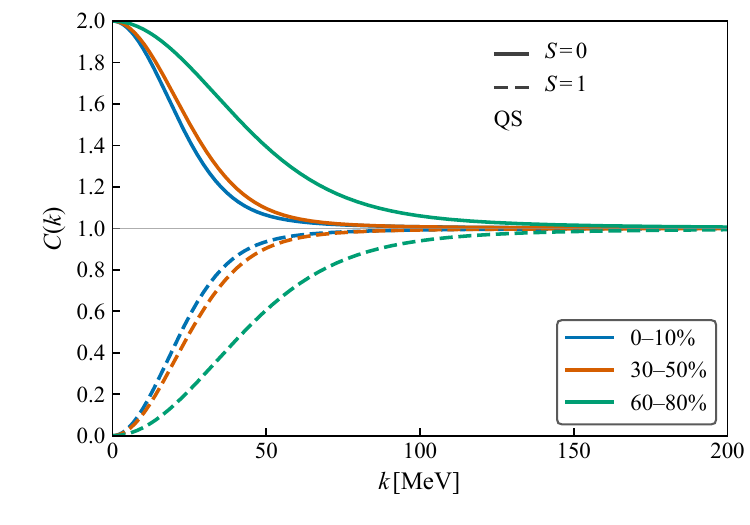}
    }
    \subfigure[]{
        \includegraphics[width=0.46\textwidth]{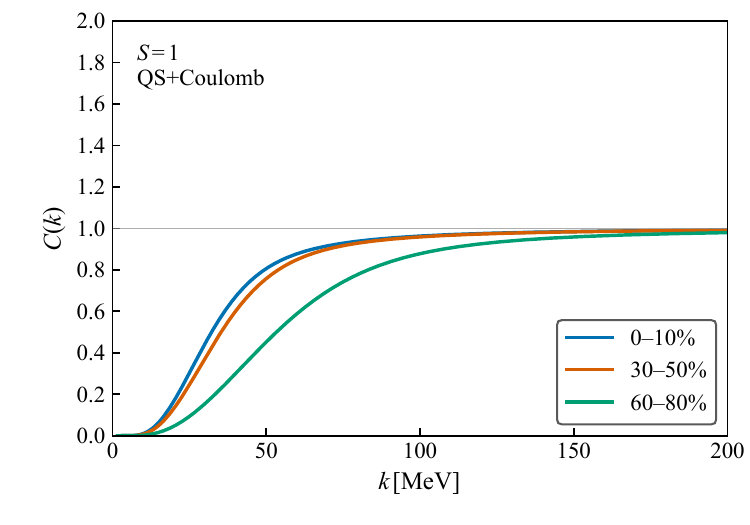}
    }
    \\[-0.2em]
    \subfigure[]{
        \includegraphics[width=0.46\textwidth]{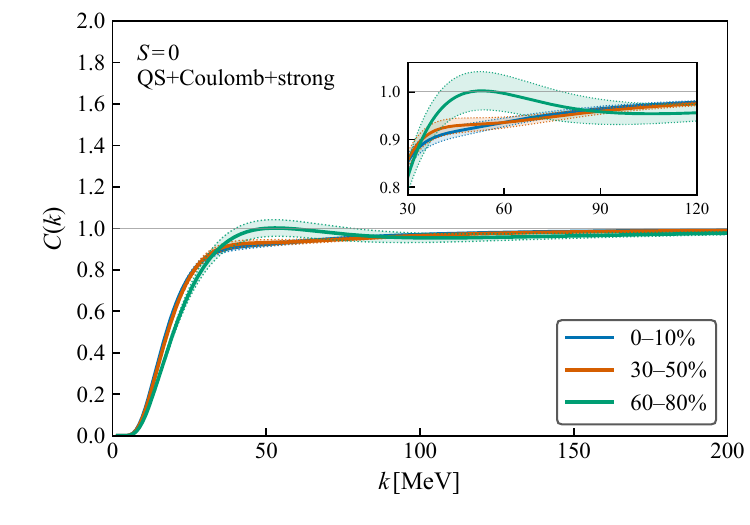}
    }
    \subfigure[]{
        \includegraphics[width=0.46\textwidth]{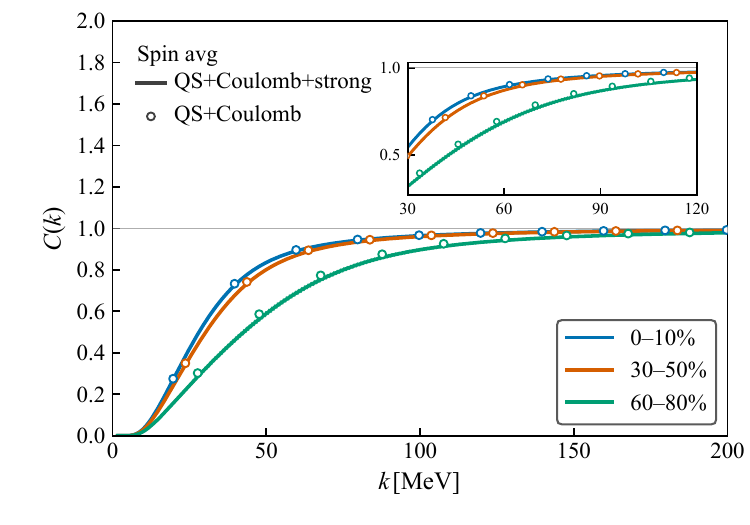}
    }
    \vspace{-0.4em}
    \caption{$\Lambda_c^+\Lambda_c^+$ momentum correlation functions for the
    0--10\%, 30--50\%, and 60--80\% centrality classes.
    Panel (a) shows the quantum-statistical reference with both Coulomb and
    strong interactions switched off: solid and dashed curves denote
    $S=0$ and $S=1$, respectively.
    Panel (b) shows the $S=1$ background evaluated with the
    antisymmetrized spatial wave function and the Coulomb interaction
    (QS+Coulomb).
    Panel (c) shows the $S=0$ EFT result including spatial
    symmetrization, the Coulomb interaction, and the strong interaction
    (QS+Coulomb+strong).
    Panel (d) shows the corresponding spin-averaged correlation (Spin avg),
    $C_{\rm avg}=\tfrac14 C_{S=0}^{\rm Coul+strong}+\tfrac34 C_{S=1}^{\rm Coul}$.
    Central parameters and fitted intervals are listed in
    Table~\ref{tab:interaction_variations}; the shaded bands in panels (c)
    and (d) propagate the combined lattice scattering-length errors.
    The central curves in panels (c) and (d) are solid.
    The spin-averaged no-strong-interaction reference,
    $C_{\rm avg}^{\rm Coul}=\tfrac14 C_{S=0}^{\rm Coul}+\tfrac34 C_{S=1}^{\rm Coul}$,
    is shown by color-matched open circles in panel (d), including its inset.
    The insets in panels (c) and (d) magnify $C(k)$ in the
    $30$--$120~\mev$ region. }
    \label{fig:lclc_components}
\end{figure}

In Fig.~\ref{fig:lclc_components}(a), both the Coulomb and the strong interaction are switched off.
For the PHSD sources, spatial symmetrization gives $C_{S=0}^{\rm QS}(0)=2$,
whereas spatial antisymmetrization gives $C_{S=1}^{\rm QS}(0)=0$; the two
components then approach unity as the exchange contribution decreases with
momentum. Figure~\ref{fig:lclc_components}(b) adds Coulomb repulsion to the $S=1$ sector only,
producing an additional low-$k$ suppression. These reference curves depend
on the source and on the indicated exchange or Coulomb effects, but not on
the strong-interaction parameters. Figure~\ref{fig:lclc_components}(c) shows the $S=0$ EFT result with
quantum-statistical symmetrization, the Coulomb interaction, and the strong
interaction included. Since the calibrated $\Lambda_c\Lambda_c$ interaction
is repulsive, it reinforces Coulomb repulsion and counteracts the
exchange-induced enhancement in the singlet channel; away from the
Coulomb-dominated threshold region, $C_{S=0}^{\rm Coul+strong}$ therefore
lies below the corresponding no-strong-interaction reference. Finally, Fig.~\ref{fig:lclc_components}(d) shows the spin-averaged correlation,
$C_{\rm avg}=\tfrac14 C_{S=0}^{\rm Coul+strong}+\tfrac34 C_{S=1}^{\rm Coul}$,
together with the no-strong-interaction reference
$C_{\rm avg}^{\rm Coul}=\tfrac14 C_{S=0}^{\rm Coul}+\tfrac34 C_{S=1}^{\rm Coul}$
shown by open circles. The pronounced low-$k$ depletion of the reference
reflects both the repulsive Coulomb interaction and the $3/4$ statistical
weight of the triplet component. Because the strong interaction enters only through the singlet weight
$1/4$, $C_{\rm avg}$ remains close to $C_{\rm avg}^{\rm Coul}$ but lies
systematically below it. The separation reaches a maximum at
intermediate momentum and decreases toward the upper end of the
displayed range. The
strong-interaction signature in $\Lambda_c^+\Lambda_c^+$ is therefore the
small but systematic displacement of $C_{\rm avg}(k)$ from
$C_{\rm avg}^{\rm Coul}(k)$, rather than the deviation of $C_{\rm avg}(k)$
from unity.

As in the $\ddsys$ case, the centrality dependence follows the source
distributions of Sec.~\ref{sec:phsd_results}: the compact 60--80\% source
enhances the low-$k$ response of both the reference and the full curves.
Accordingly, the $\Lambda_c^+\Lambda_c^+$ correlation is most pronounced in
peripheral collisions, where the small source maximizes the low-momentum
signal. Nevertheless, because the reference itself is strongly suppressed at
low $k$ by Coulomb repulsion and by the triplet statistical weight, the
strong-interaction effect should be extracted from the difference
$C_{\rm avg}-C_{\rm avg}^{\rm Coul}$, or equivalently from the ratio
$C_{\rm avg}/C_{\rm avg}^{\rm Coul}$, rather than from the absolute value of
$C_{\rm avg}$ alone. The shaded bands of the $S=0$ component and of the
spin-averaged result propagate the lattice scattering-length uncertainty,
whereas the $S=1$ background is unchanged; the theory uncertainty of the
strong-interaction signature is thus itself weighted by the factor $1/4$.

A qualitative contrast with the neutral $\Lambda\Lambda$ system clarifies why
the charged $\Lambda_c^+\Lambda_c^+$ signal should not be read from the
absolute low-$k$ normalization. In a Coulomb-free and strong-interaction-free
reference, the symmetric $S=0$ wave function gives $C_{S=0}^{\rm QS}(0)=2$,
while the antisymmetric $S=1$ component vanishes at threshold; with the
$1/4$ and $3/4$ spin weights this would give $C_{\rm avg}^{\rm QS}(0)=1/2$.
For charged $\Lambda_c^+\Lambda_c^+$, however, Coulomb repulsion strongly
depletes the correlation at low $k$ in both spin sectors. The observed
low-momentum suppression is therefore dominated by the source and by
Coulomb/exchange effects, and the strong-interaction information is carried
only by the small displacement from $C_{\rm avg}^{\rm Coul}$. This comparison
is qualitative and does not assume identical $\Lambda\Lambda$ and
$\Lambda_c^+\Lambda_c^+$ sources or interactions.
\FloatBarrier

\section{Summary}\label{sec:summary}

In summary, we have constructed lattice-guided descriptions of the near-threshold
\ddsys\ and \lclc\ interactions—an OBE potential and an EFT contact
interaction for \ddsys, and an EFT contact interaction for \lclc—and folded the
resulting scattering wave functions with species-dependent, non-Gaussian PHSD
sources to obtain femtoscopic correlation predictions for $D^+D^+$, $D^0D^0$,
$D^+D^0$, and $\Lambda_c^+\Lambda_c^+$ pairs in Pb--Pb collisions at
$\sqrt{s_{\rm NN}}=5.02~\tev$. The scattering lengths, which are the best
constrained lattice quantities, fix the threshold strengths of both interaction
descriptions, while the effective ranges and finite-energy phase shifts serve as
independent, unfitted diagnostics; for \lclc, the fixed-coupling OBE setup cannot
reproduce the scattering length and phase shifts simultaneously and is therefore not
retained. The PHSD sources become more compact from central to peripheral
collisions and retain non-Gaussian shape information, which strengthens the
low-momentum signal, most notably for the peripheral \lclc\ channel. It should be emphasized that the resulting correlation functions are conditional benchmarks: they are obtained with the lattice-calibrated threshold strengths and the PHSD source distributions held fixed, and the shaded bands reflect only the propagated lattice scattering-length uncertainty, not the source-fit, regulator, coupling, or pion-mass-extrapolation uncertainties discussed above.

The four physical channels separate the ingredients that enter the correlation
function. Among the \ddsys\ pairs, $D^+D^+$ contains Coulomb, quantum-statistical,
and strong contributions; $D^0D^0$ isolates the interplay of Bose--Einstein
symmetrization and the strong interaction; and $D^{+}D^{0}$, free of both Coulomb and identical-particle effects, provides the cleanest and most transparent access to the $DD$ strong interaction, albeit with a signal of only a few percent because the calibrated scattering length is small; resolving this signal will require the statistical precision of the upgraded ALICE heavy-ion program, and the absence of competing backgrounds in this channel makes it the most favorable case for such a measurement.  For $\Lambda_c^+\Lambda_c^+$, the
spin-triplet component acts as a fixed background carrying no strong $S$-wave
interaction, so the strong-interaction effect appears in the spin-averaged
observable as a small, systematic shift—weighted by the factor $1/4$ of the
singlet—relative to the no-strong-interaction reference. This shift is computed with the $\Lambda_{c}\Lambda_{c}$ scattering length calibrated at a nonphysical pion mass; while a chiral-EFT extrapolation indicates that the interaction remains repulsive at the physical point~\cite{Liu:2026yhn}, the magnitude of the shift at the physical pion mass is not quantitatively predicted here.  Because both \ddsys\
models are calibrated to the same lattice scattering length, their correlation
functions remain close in the low-momentum region where the signal is largest;
the residual separation at finite momentum carries sensitivity to the effective
range and to interaction structure beyond the scattering length.

Looking forward, several developments would sharpen these benchmark estimates and their experimental use.  On the theory side, refined lattice QCD results at the physical
pion mass for \lclc, and extended finite-volume spectra for both systems, would
reduce the dominant input uncertainty propagated here; including coupled-channel
effects, odd-partial-wave interactions, and higher-order EFT terms would allow the
finite-momentum region to be used as a genuine discriminator between interaction
shapes rather than a residual model dependence. The dependence of the \lclc\
calibration at the nonphysical lattice pion mass should also be quantified. On the
experimental side, the $D^+D^0$ correlation in Pb--Pb collisions is the most
direct observable for the repulsive \ddsys\ interaction, while
$\Lambda_c^+\Lambda_c^+$ should be measured preferentially in peripheral events,
where the compact source maximizes the low-momentum signal; a joint analysis of
all four channels, together with the model-independent Coulomb and
quantum-statistical references constructed here, would allow the strong,
Coulomb, and statistical contributions to be disentangled without external
input. Such measurements, within reach of the high-statistics heavy-ion program of the upgraded ALICE experiment, would test these benchmark estimates and, once the source and interaction-shape uncertainties are quantified, could turn these reference channels into quantitative constraints on near-threshold charmed-hadron interactions.

\section{Acknowledgments}

This work is supported by the National Natural Science Foundation of China under Grant  No.~12622505, No.~12575086, and  No.~W2543006. Qi-Wei Yuan acknowledges support from  the
Fundamental Research Funds for the Central Universities under No.~lzujbky-2023-stlt01.

\renewcommand{\bibfont}{\small\setlength{\baselineskip}{0.95\baselineskip}}
\bibliography{reference}

@article{Shi:2025ogt,
    author = "Shi, Pan-Pan and Guo, Feng-Kun and Liu, Chuan and Liu, Liuming and Sun, Peng and Wu, Jia-Jun and Xing, Hanyang",
    title = "{Low-energy DD scattering in lattice QCD}",
    eprint = "2502.07438",
    archivePrefix = "arXiv",
    primaryClass = "hep-lat",
    doi = "10.1103/6xsx-ldwc",
    journal = "Phys. Rev. D",
    volume = "114",
    number = "3",
    pages = "034505",
    year = "2026"
}

@article{Epelbaum:2014sza,
    author = "Epelbaum, E. and Krebs, H. and Mei{\ss}ner, U. G.",
    title = "{Precision nucleon-nucleon potential at fifth order in the chiral expansion}",
    eprint = "1412.4623",
    archivePrefix = "arXiv",
    primaryClass = "nucl-th",
    doi = "10.1103/PhysRevLett.115.122301",
    journal = "Phys. Rev. Lett.",
    volume = "115",
    number = "12",
    pages = "122301",
    year = "2015"
}

@article{Aoki:2012tk,
    author = "Aoki, Sinya and Doi, Takumi and Hatsuda, Tetsuo and Ikeda, Yoichi and Inoue, Takashi and Ishii, Noriyoshi and Murano, Keiko and Nemura, Hidekatsu and Sasaki, Kenji",
    collaboration = "HAL QCD",
    title = "{Lattice QCD approach to Nuclear Physics}",
    eprint = "1206.5088",
    archivePrefix = "arXiv",
    primaryClass = "hep-lat",
    doi = "10.1093/ptep/pts010",
    journal = "PTEP",
    volume = "2012",
    pages = "01A105",
    year = "2012"
}

@article{Luscher:1991cf,
    author = "Luscher, Martin",
    title = "{Signatures of unstable particles in finite volume}",
    reportNumber = "DESY-91-052",
    doi = "10.1016/0550-3213(91)90584-K",
    journal = "Nucl. Phys. B",
    volume = "364",
    pages = "237--251",
    year = "1991"
}

@article{BMW:2008jgk,
    author = "Durr, S. and others",
    collaboration = "BMW",
    title = "{Ab-Initio Determination of Light Hadron Masses}",
    eprint = "0906.3599",
    archivePrefix = "arXiv",
    primaryClass = "hep-lat",
    doi = "10.1126/science.1163233",
    journal = "Science",
    volume = "322",
    pages = "1224--1227",
    year = "2008"
}

@article{Pan:2025vyg,
    author = "Pan, Ya-Wen and Liu, Ming-Zhu and Geng, Li-Sheng",
    title = "{Probing the Three-Body Force in Hadronic Systems with Specific Charge Parity}",
    eprint = "2512.01468",
    archivePrefix = "arXiv",
    primaryClass = "nucl-th",
    doi = "10.1103/vl2y-8xdn",
    journal = "Phys. Rev. Lett.",
    volume = "136",
    number = "15",
    pages = "151901",
    year = "2026"
}

@article{Li:1996yn,
    author = "Li, Xue-Qian and Bugg, David V. and Zou, Bing-Song",
    title = "{A Possible explanation of the 'rho pi puzzle' in J / psi, psi-prime decays}",
    reportNumber = "RAL-TR-96-062",
    doi = "10.1103/PhysRevD.55.1421",
    journal = "Phys. Rev. D",
    volume = "55",
    pages = "1421--1424",
    year = "1997"
}

@article{Chen:2016qju,
    author = "Chen, Hua-Xing and Chen, Wei and Liu, Xiang and Zhu, Shi-Lin",
    title = "{The hidden-charm pentaquark and tetraquark states}",
    eprint = "1601.02092",
    archivePrefix = "arXiv",
    primaryClass = "hep-ph",
    doi = "10.1016/j.physrep.2016.05.004",
    journal = "Phys. Rept.",
    volume = "639",
    pages = "1--121",
    year = "2016"
}

@article{Liu:2024uxn,
    author = "Liu, Ming-Zhu and Pan, Ya-Wen and Liu, Zhi-Wei and Wu, Tian-Wei and Lu, Jun-Xu and Geng, Li-Sheng",
    title = "{Three ways to decipher the nature of exotic hadrons: Multiplets, three-body hadronic molecules, and correlation functions}",
    eprint = "2404.06399",
    archivePrefix = "arXiv",
    primaryClass = "hep-ph",
    doi = "10.1016/j.physrep.2024.12.001",
    journal = "Phys. Rept.",
    volume = "1108",
    pages = "1--108",
    year = "2025"
}

@article{Xing:2025uai,
    author = "Xing, Hanyang and Geng, Yiqi and Liu, Chuan and Liu, Liuming and Sun, Peng and Wu, Jiajun and Yan, Zhicheng and Zhu, Ruilin",
    title = "{Lattice QCD study of {\ensuremath{\Lambda}}$_{c}${\ensuremath{\Lambda}}$_{c}$ scattering*}",
    eprint = "2502.05546",
    archivePrefix = "arXiv",
    primaryClass = "hep-lat",
    doi = "10.1088/1674-1137/adc11f",
    journal = "Chin. Phys. C",
    volume = "49",
    number = "6",
    pages = "063107",
    year = "2025"
}

@article{Luscher:1986pf,
    author = "Luscher, M.",
    title = "{Volume Dependence of the Energy Spectrum in Massive Quantum Field Theories. 2. Scattering States}",
    reportNumber = "DESY-86-034",
    doi = "10.1007/BF01211097",
    journal = "Commun. Math. Phys.",
    volume = "105",
    pages = "153--188",
    year = "1986"
}

@article{Luscher:1990ux,
    author = "Luscher, Martin",
    title = "{Two particle states on a torus and their relation to the scattering matrix}",
    reportNumber = "DESY-90-131",
    doi = "10.1016/0550-3213(91)90366-6",
    journal = "Nucl. Phys. B",
    volume = "354",
    pages = "531--578",
    year = "1991"
}

@article{Liu:2019zvb,
    author = "Liu, Ming-Zhu and Wu, Tian-Wei and S{\'a}nchez S{\'a}nchez, Mario and Valderrama, Manuel Pavon and Geng, Li-Sheng and Xie, Ju-Jun",
    title = "{Spin-parities of the $P_c(4440)$ and $P_c(4457)$ in the one-boson-exchange model}",
    eprint = "1907.06093",
    archivePrefix = "arXiv",
    primaryClass = "hep-ph",
    doi = "10.1103/PhysRevD.103.054004",
    journal = "Phys. Rev. D",
    volume = "103",
    number = "5",
    pages = "054004",
    year = "2021"
}

@article{Lee:2011rka,
    author = "Lee, Ning and Luo, Zhi-Gang and Chen, Xiao-Lin and Zhu, Shi-Lin",
    title = "{Possible Deuteron-like Molecular States Composed of Heavy Baryons}",
    eprint = "1104.4257",
    archivePrefix = "arXiv",
    primaryClass = "hep-ph",
    doi = "10.1103/PhysRevD.84.014031",
    journal = "Phys. Rev. D",
    volume = "84",
    pages = "014031",
    year = "2011"
}

@article{Lu:2017dvm,
    author = "Lu, Jun-Xu and Geng, Li-Sheng and Valderrama, Manuel Pavon",
    title = "{Heavy baryon-antibaryon molecules in effective field theory}",
    eprint = "1706.02588",
    archivePrefix = "arXiv",
    primaryClass = "hep-ph",
    doi = "10.1103/PhysRevD.99.074026",
    journal = "Phys. Rev. D",
    volume = "99",
    number = "7",
    pages = "074026",
    year = "2019"
}

@article{Koonin:1977fh,
    author = "Koonin, S. E.",
    title = "{Proton Pictures of High-Energy Nuclear Collisions}",
    doi = "10.1016/0370-2693(77)90340-9",
    journal = "Phys. Lett. B",
    volume = "70",
    pages = "43--47",
    year = "1977"
}

@article{Pratt:1986cc,
    author = "Pratt, S.",
    title = "{Pion Interferometry of Quark-Gluon Plasma}",
    doi = "10.1103/PhysRevD.33.1314",
    journal = "Phys. Rev. D",
    volume = "33",
    pages = "1314--1327",
    year = "1986"
}

@article{Cassing:2008sv,
    author = "Cassing, W. and Bratkovskaya, E. L.",
    title = "{Parton transport and hadronization from the dynamical quasiparticle point of view}",
    eprint = "0808.0022",
    archivePrefix = "arXiv",
    primaryClass = "hep-ph",
    doi = "10.1103/PhysRevC.78.034919",
    journal = "Phys. Rev. C",
    volume = "78",
    pages = "034919",
    year = "2008"
}

@article{Bratkovskaya:2011wp,
    author = "Bratkovskaya, E. L. and Cassing, W. and Konchakovski, V. P. and Linnyk, O.",
    title = "{Parton-Hadron-String Dynamics at Relativistic Collider Energies}",
    eprint = "1101.5793",
    archivePrefix = "arXiv",
    primaryClass = "nucl-th",
    doi = "10.1016/j.nuclphysa.2011.03.003",
    journal = "Nucl. Phys. A",
    volume = "856",
    pages = "162--182",
    year = "2011"
}

@article{LHCb:2012aiv,
    author = "Aaij, R and others",
    collaboration = "LHCb",
    title = "{Observation of double charm production involving open charm in pp collisions at $\sqrt{s}$ = 7 TeV}",
    eprint = "1205.0975",
    archivePrefix = "arXiv",
    primaryClass = "hep-ex",
    reportNumber = "CERN-PH-EP-2012-109, LHCB-PAPER-2012-003",
    doi = "10.1007/JHEP06(2012)141",
    journal = "JHEP",
    volume = "06",
    pages = "141",
    year = "2012",
    note = "[Addendum: JHEP 03, 108 (2014)]"
}

@article{Song:2018xca,
    author = "Song, Taesoo and Cassing, Wolfgang and Moreau, Pierre and Bratkovskaya, Elena",
    title = "{Open charm and dileptons from relativistic heavy-ion collisions}",
    eprint = "1803.02698",
    archivePrefix = "arXiv",
    primaryClass = "nucl-th",
    doi = "10.1103/PhysRevC.97.064907",
    journal = "Phys. Rev. C",
    volume = "97",
    number = "6",
    pages = "064907",
    year = "2018"
}

@article{Lisa:2005dd,
    author = "Lisa, Michael Annan and Pratt, Scott and Soltz, Ron and Wiedemann, Urs",
    title = "{Femtoscopy in relativistic heavy ion collisions}",
    eprint = "nucl-ex/0505014",
    archivePrefix = "arXiv",
    doi = "10.1146/annurev.nucl.55.090704.151533",
    journal = "Ann. Rev. Nucl. Part. Sci.",
    volume = "55",
    pages = "357--402",
    year = "2005"
}

@article{Fabbietti:2020bfg,
    author = "Fabbietti, L. and Mantovani Sarti, V. and Vazquez Doce, O.",
    title = "{Study of the Strong Interaction Among Hadrons with Correlations at the LHC}",
    eprint = "2012.09806",
    archivePrefix = "arXiv",
    primaryClass = "nucl-ex",
    doi = "10.1146/annurev-nucl-102419-034438",
    journal = "Ann. Rev. Nucl. Part. Sci.",
    volume = "71",
    pages = "377--402",
    year = "2021"
}

@article{ALICE:2024bhk,
    author = "Acharya, Shreyasi and others",
    collaboration = "ALICE",
    title = "{Studying the interaction between charm and light-flavor mesons}",
    eprint = "2401.13541",
    archivePrefix = "arXiv",
    primaryClass = "nucl-ex",
    reportNumber = "CERN-EP-2024-013",
    doi = "10.1103/PhysRevD.110.032004",
    journal = "Phys. Rev. D",
    volume = "110",
    number = "3",
    pages = "032004",
    year = "2024"
}

@article{Haidenbauer:2020kwo,
    author = "Haidenbauer, J. and Krein, G. and Peixoto, T. C.",
    title = "{Femtoscopic correlations and the $\Lambda_c N$ interaction}",
    eprint = "2004.08136",
    archivePrefix = "arXiv",
    primaryClass = "nucl-th",
    doi = "10.1140/epja/s10050-020-00190-0",
    journal = "Eur. Phys. J. A",
    volume = "56",
    number = "7",
    pages = "184",
    year = "2020"
}

@article{Torres-Rincon:2023qll,
    author = "Torres-Rincon, Juan M. and Ramos, {\`A}ngels and Tolos, Laura",
    title = "{Femtoscopy of D mesons and light mesons upon unitarized effective field theories}",
    eprint = "2307.02102",
    archivePrefix = "arXiv",
    primaryClass = "hep-ph",
    doi = "10.1103/PhysRevD.108.096008",
    journal = "Phys. Rev. D",
    volume = "108",
    number = "9",
    pages = "096008",
    year = "2023"
}

@article{Mihaylov:2018rva,
    author = "Mihaylov, D. L. and Mantovani Sarti, V. and Arnold, O. W. and Fabbietti, L. and Hohlweger, B. and Mathis, A. M.",
    title = {{A femtoscopic Correlation Analysis Tool using the Schr{\"o}dinger equation (CATS)}},
    eprint = "1802.08481",
    archivePrefix = "arXiv",
    primaryClass = "hep-ph",
    doi = "10.1140/epjc/s10052-018-5859-0",
    journal = "Eur. Phys. J. C",
    volume = "78",
    number = "5",
    pages = "394",
    year = "2018"
}

@article{ALICE:2018ysd,
    author = "Acharya, Shreyasi and others",
    collaboration = "ALICE",
    title = "{p-p, p-$\Lambda$ and $\Lambda$-$\Lambda$ correlations studied via femtoscopy in pp reactions at $\sqrt{s}$ = 7 TeV}",
    eprint = "1805.12455",
    archivePrefix = "arXiv",
    primaryClass = "nucl-ex",
    reportNumber = "CERN-EP-2018-150",
    doi = "10.1103/PhysRevC.99.024001",
    journal = "Phys. Rev. C",
    volume = "99",
    number = "2",
    pages = "024001",
    year = "2019"
}

@article{ALICE:2019eol,
    author = "Acharya, Shreyasi and others",
    collaboration = "ALICE",
    title = "{Study of the $\Lambda$-$\Lambda$ interaction with femtoscopy correlations in pp and p-Pb collisions at the LHC}",
    eprint = "1905.07209",
    archivePrefix = "arXiv",
    primaryClass = "nucl-ex",
    reportNumber = "CERN-EP-2019-096",
    doi = "10.1016/j.physletb.2019.134822",
    journal = "Phys. Lett. B",
    volume = "797",
    pages = "134822",
    year = "2019"
}

@article{Molina:2025lzw,
    author = "Molina, R. and Oset, E.",
    title = "{Determination of off-shell ambiguities in correlation functions: Strategies to minimize them}",
    eprint = "2506.03669",
    archivePrefix = "arXiv",
    primaryClass = "hep-ph",
    doi = "10.1103/rst4-rkmm",
    journal = "Phys. Rev. D",
    volume = "112",
    number = "9",
    pages = "096006",
    year = "2025"
}

@article{Dai:2026fkg,
    author = "Dai, Xinchen and Jia, Sen and Nefediev, Alexey and Nieves, Juan and Shen, Chengping and Zhang, Liming",
    title = "{Exotic hadrons associated with b-quark}",
    eprint = "2603.09315",
    archivePrefix = "arXiv",
    primaryClass = "hep-ph",
    doi = "10.1016/j.physrep.2026.06.002",
    journal = "Phys. Rept.",
    volume = "1191",
    pages = "1--62",
    year = "2026"
}

@article{Gobel:2025afq,
    author = {G{\"o}bel, Matthias and Kievsky, Alejandro},
    title = "{Nucleon-nucleon correlation functions from different interactions in comparison}",
    eprint = "2505.13433",
    archivePrefix = "arXiv",
    primaryClass = "nucl-th",
    doi = "10.1016/j.physletb.2025.139835",
    journal = "Phys. Lett. B",
    volume = "869",
    pages = "139835",
    year = "2025"
}

@article{Zhao:2026bhh,
    author = "Zhao, Jiaxing and Song, Taesoo and Bratkovskaya, Elena and Aichelin, Joerg",
    title = "{Accessing Exotic Hadronic States via Charmed-Meson Femtoscopy in Relativistic Heavy-Ion Collisions}",
    eprint = "2605.31527",
    archivePrefix = "arXiv",
    primaryClass = "nucl-th",
    month = "5",
    year = "2026"
}

@article{ALICE:2018hbc,
    author = "Acharya, Shreyasi and others",
    collaboration = "ALICE",
    title = "{$\Lambda_\mathrm{c}^+$ production in Pb-Pb collisions at $\sqrt{s_{\rm NN}} = 5.02$ TeV}",
    eprint = "1809.10922",
    archivePrefix = "arXiv",
    primaryClass = "nucl-ex",
    reportNumber = "CERN-EP-2018-261",
    doi = "10.1016/j.physletb.2019.04.046",
    journal = "Phys. Lett. B",
    volume = "793",
    pages = "212--223",
    year = "2019"
}

@article{ALICE:2018lyv,
    author = "Acharya, S. and others",
    collaboration = "ALICE",
    title = "{Measurement of D$^{0}$, D$^{+}$, D$^{*+}$ and D$_{s}^{+}$ production in Pb-Pb collisions at $ \sqrt{{\mathrm{s}}_{\mathrm{NN}}}=5.02 $ TeV}",
    eprint = "1804.09083",
    archivePrefix = "arXiv",
    primaryClass = "nucl-ex",
    reportNumber = "CERN-EP-2018-066",
    doi = "10.1007/JHEP10(2018)174",
    journal = "JHEP",
    volume = "10",
    pages = "174",
    year = "2018"
}

@article{Bai:2026syt,
    author = "Bai, Xiaozhi and others",
    title = "{Hyperon-Nucleon Spectrometer}",
    eprint = "2606.06553",
    archivePrefix = "arXiv",
    primaryClass = "physics.ins-det",
    month = "6",
    year = "2026"
}

@article{Lu:2025syk,
    author = "Lu, Jun-Xu and Xiao, Yang and Liu, Zhi-Wei and Geng, Li-Sheng",
    title = "{Relativistic chiral nuclear forces: Status and prospects}",
    eprint = "2501.17185",
    archivePrefix = "arXiv",
    primaryClass = "nucl-th",
    doi = "10.1142/S0218301325430074",
    journal = "Int. J. Mod. Phys. E",
    volume = "34",
    number = "11",
    pages = "2543007",
    year = "2025"
}

@article{ALICE:2019hdt,
    author = "Acharya, Shreyasi and others",
    collaboration = "ALICE",
    title = "{First Observation of an Attractive Interaction between a Proton and a Cascade Baryon}",
    eprint = "1904.12198",
    archivePrefix = "arXiv",
    primaryClass = "nucl-ex",
    reportNumber = "CERN-EP-2019-080",
    doi = "10.1103/PhysRevLett.123.112002",
    journal = "Phys. Rev. Lett.",
    volume = "123",
    number = "11",
    pages = "112002",
    year = "2019"
}

@article{ALICE:2019buq,
    author = "Acharya, Shreyasi and others",
    collaboration = "ALICE",
    title = "{Investigation of the p{\textendash}{\ensuremath{\Sigma}}0 interaction via femtoscopy in pp collisions}",
    eprint = "1910.14407",
    archivePrefix = "arXiv",
    primaryClass = "nucl-ex",
    reportNumber = "CERN-EP-2019-232",
    doi = "10.1016/j.physletb.2020.135419",
    journal = "Phys. Lett. B",
    volume = "805",
    pages = "135419",
    year = "2020"
}

@article{PitangaLachini:2025pxr,
    author = "Pitanga Lachini, Nelson and Thomas, Christopher E. and Wilson, David J.",
    collaboration = "Hadron Spectrum",
    title = "{Coupled-channel scattering of DD, DD*, and D*D* in isospin-1 from lattice QCD}",
    eprint = "2505.01363",
    archivePrefix = "arXiv",
    primaryClass = "hep-lat",
    doi = "10.1103/xz5r-p533",
    journal = "Phys. Rev. D",
    volume = "112",
    number = "11",
    pages = "114511",
    year = "2025"
}

@article{ALICE:2015hvw,
    author = "Adam, Jaroslav and others",
    collaboration = "ALICE",
    title = "{One-dimensional pion, kaon, and proton femtoscopy in Pb-Pb collisions at $\sqrt{s_{\rm {NN}}}$ =2.76 TeV}",
    eprint = "1506.07884",
    archivePrefix = "arXiv",
    primaryClass = "nucl-ex",
    reportNumber = "CERN-PH-EP-2015-106",
    doi = "10.1103/PhysRevC.92.054908",
    journal = "Phys. Rev. C",
    volume = "92",
    number = "5",
    pages = "054908",
    year = "2015"
}

@article{Yalikun:2021dpk,
    author = "Yalikun, Nijiati and Zou, Bing-Song",
    title = "{Anticharmed strange pentaquarks from the one-boson-exchange model}",
    eprint = "2112.06426",
    archivePrefix = "arXiv",
    primaryClass = "hep-ph",
    doi = "10.1103/PhysRevD.105.094026",
    journal = "Phys. Rev. D",
    volume = "105",
    number = "9",
    pages = "094026",
    year = "2022"
}

@article{Shen:2025qpj,
    author = "Shen, Yi-bo and Liu, Zhi-Wei and Lu, Jun-Xu and Liu, Ming-Zhu and Geng, Li-Sheng",
    title = "{Probing the structure of the $D_{s 0}^*(2317)$ and $X(3872)$ states through correlation functions}",
    eprint = "2506.23476",
    archivePrefix = "arXiv",
    primaryClass = "hep-ph",
    doi = "10.1016/j.physletb.2026.140535",
    journal = "Phys. Lett. B",
    volume = "878",
    pages = "140535",
    year = "2026"
}

@article{ALICE:2019gcn,
    author = "Acharya, Shreyasi and others",
    collaboration = "ALICE",
    title = "{Scattering studies with low-energy kaon-proton femtoscopy in proton-proton collisions at the LHC}",
    eprint = "1905.13470",
    archivePrefix = "arXiv",
    primaryClass = "nucl-ex",
    reportNumber = "CERN-EP-2019-118",
    doi = "10.1103/PhysRevLett.124.092301",
    journal = "Phys. Rev. Lett.",
    volume = "124",
    number = "9",
    pages = "092301",
    year = "2020"
}

@article{ALICE:2020mfd,
    author = "Acharya, Shreyasi and others",
    collaboration = "ALICE",
    title = "{Unveiling the strong interaction among hadrons at the LHC}",
    eprint = "2005.11495",
    archivePrefix = "arXiv",
    primaryClass = "nucl-ex",
    reportNumber = "CERN-EP-2020-091",
    doi = "10.1038/s41586-020-3001-6",
    journal = "Nature",
    volume = "588",
    pages = "232--238",
    year = "2020",
    note = "[Erratum: Nature 590, E13 (2021)]"
}

@article{ALICE:2022enj,
    author = "Acharya, Shreyasi and others",
    collaboration = "ALICE",
    title = "{First study of the two-body scattering involving charm hadrons}",
    eprint = "2201.05352",
    archivePrefix = "arXiv",
    primaryClass = "nucl-ex",
    reportNumber = "CERN-EP-2022-006",
    doi = "10.1103/PhysRevD.106.052010",
    journal = "Phys. Rev. D",
    volume = "106",
    number = "5",
    pages = "052010",
    year = "2022"
}

@article{Kamiya:2022thy,
    author = "Kamiya, Yuki and Hyodo, Tetsuo and Ohnishi, Akira",
    title = "{Femtoscopic study on $DD^*$ and $D\bar{D}^*$ interactions for $T_{cc}$ and X(3872)}",
    eprint = "2203.13814",
    archivePrefix = "arXiv",
    primaryClass = "hep-ph",
    reportNumber = "YITP-22-26, RIKEN-iTHEMS-Report-22",
    doi = "10.1140/epja/s10050-022-00782-y",
    journal = "Eur. Phys. J. A",
    volume = "58",
    number = "7",
    pages = "131",
    year = "2022"
}

@article{Vidana:2023olz,
    author = "Vidana, I. and Feijoo, A. and Albaladejo, M. and Nieves, J. and Oset, E.",
    title = "{Femtoscopic correlation function for the Tcc(3875)+ state}",
    eprint = "2303.06079",
    archivePrefix = "arXiv",
    primaryClass = "hep-ph",
    doi = "10.1016/j.physletb.2023.138201",
    journal = "Phys. Lett. B",
    volume = "846",
    pages = "138201",
    year = "2023"
}

@article{Morita:2019rph,
    author = "Morita, Kenji and Gongyo, Shinya and Hatsuda, Tetsuo and Hyodo, Tetsuo and Kamiya, Yuki and Ohnishi, Akira",
    title = "{Probing $\Omega\Omega$ and $p\Omega$ dibaryons with femtoscopic correlations in relativistic heavy-ion collisions}",
    eprint = "1908.05414",
    archivePrefix = "arXiv",
    primaryClass = "nucl-th",
    reportNumber = "YITP-19-79, RIKEN-QHP-423, RIKEN-iTHEMS-Report-19",
    doi = "10.1103/PhysRevC.101.015201",
    journal = "Phys. Rev. C",
    volume = "101",
    number = "1",
    pages = "015201",
    year = "2020"
}

@article{Kamiya:2019uiw,
    author = "Kamiya, Yuki and Hyodo, Tetsuo and Morita, Kenji and Ohnishi, Akira and Weise, Wolfram",
    title = "{$K^-p$ Correlation Function from High-Energy Nuclear Collisions and Chiral SU(3) Dynamics}",
    eprint = "1911.01041",
    archivePrefix = "arXiv",
    primaryClass = "nucl-th",
    doi = "10.1103/PhysRevLett.124.132501",
    journal = "Phys. Rev. Lett.",
    volume = "124",
    number = "13",
    pages = "132501",
    year = "2020"
}

@article{Kamiya:2021hdb,
    author = "Kamiya, Y. and Sasaki, K. and Fukui, T. and Hyodo, T. and Morita, K. and Ogata, K. and Ohnishi, A. and Hatsuda, T.",
    title = "{Femtoscopic study of coupled-channels $N \Xi$ and $\Lambda\Lambda$ interactions}",
    eprint = "2108.09644",
    archivePrefix = "arXiv",
    primaryClass = "hep-ph",
    reportNumber = "YITP-21-79, RIKEN-iTHEMS-Report-21, NITEP 116",
    doi = "10.1103/PhysRevC.105.014915",
    journal = "Phys. Rev. C",
    volume = "105",
    number = "1",
    pages = "014915",
    year = "2022"
}

@article{ALICE:2023sjd,
    author = "Acharya, Shreyasi and others",
    collaboration = "ALICE",
    title = "{Common femtoscopic hadron-emission source in pp collisions at the LHC}",
    eprint = "2311.14527",
    archivePrefix = "arXiv",
    primaryClass = "hep-ph",
    reportNumber = "CERN-EP-2023-267",
    doi = "10.1140/epjc/s10052-025-13793-y",
    journal = "Eur. Phys. J. C",
    volume = "85",
    number = "2",
    pages = "198",
    year = "2025",
    note = "[Erratum: Eur.Phys.J.C 86, 12 (2026)]"
}

@article{Epelbaum:2025aan,
    author = "Epelbaum, Evgeny and Heihoff, Sven and Mei{\ss}ner, Ulf-G. and Tscherwon, Alexander",
    title = "{Can the Strong Interactions between Hadrons Be Determined Using Femtoscopy?}",
    eprint = "2504.08631",
    archivePrefix = "arXiv",
    primaryClass = "nucl-th",
    doi = "10.1103/tfsb-wlsd",
    journal = "Phys. Rev. Lett.",
    volume = "136",
    number = "21",
    pages = "212301",
    year = "2026"
}

@article{Xiong:2025bmd,
    author = "Xiong, Ao-Sheng and Yuan, Qi-Wei and Liu, Ming-Zhu and Yu, Fu-Sheng and Liu, Zhi-Wei and Geng, Li-Sheng",
    title = "{Solving the inverse source problem in femtoscopy with a toy model*}",
    eprint = "2512.06904",
    archivePrefix = "arXiv",
    primaryClass = "hep-ph",
    doi = "10.1088/1674-1137/ae6310",
    journal = "Chin. Phys. C",
    volume = "50",
    number = "8",
    pages = "081001",
    year = "2026"
}

@article{Liu:2026yhn,
    author = "Liu, Zhe and Xu, Hao and Liu, Zhan-Wei and Liu, Xiang",
    title = "{Scattering of {\ensuremath{\Lambda}}c{\ensuremath{\Lambda}}c and {\ensuremath{\Lambda}}c{\ensuremath{\Lambda}}{\textasciimacron}c in chiral effective field theory}",
    eprint = "2603.09404",
    archivePrefix = "arXiv",
    primaryClass = "hep-ph",
    doi = "10.1103/7r33-df7z",
    journal = "Phys. Rev. D",
    volume = "114",
    number = "5",
    pages = "054042",
    year = "2026"
}

@article{ParticleDataGroup:2024cfk,
    author = "Navas, S. and others",
    collaboration = "Particle Data Group",
    title = "{Review of particle physics}",
    doi = "10.1103/PhysRevD.110.030001",
    journal = "Phys. Rev. D",
    volume = "110",
    number = "3",
    pages = "030001",
    year = "2024"
}

@article{Bai:2026atm,
    author = "Bai, Zi-Yue and Chen, Dian-Yong and Huang, Qi and Liu, Xiang and Luo, Si-Qiang and Wang, Jun-Zhang",
    title = "{Unquenched charmonium and beyond}",
    eprint = "2602.19887",
    archivePrefix = "arXiv",
    primaryClass = "hep-ph",
    doi = "10.1016/j.physrep.2026.08.002",
    journal = "Phys. Rept.",
    volume = "1204",
    pages = "1--162",
    year = "2027"
}

@article{Liu:2023uly,
    author = "Liu, Zhi-Wei and Lu, Jun-Xu and Geng, Li-Sheng",
    title = "{Study of the DK interaction with femtoscopic correlation functions}",
    eprint = "2302.01046",
    archivePrefix = "arXiv",
    primaryClass = "hep-ph",
    doi = "10.1103/PhysRevD.107.074019",
    journal = "Phys. Rev. D",
    volume = "107",
    number = "7",
    pages = "074019",
    year = "2023"
}

@article{Oset:2016lyh,
    author = "Oset, Eulogio and others",
    title = "{Weak decays of heavy hadrons into dynamically generated resonances}",
    eprint = "1601.03972",
    archivePrefix = "arXiv",
    primaryClass = "hep-ph",
    doi = "10.1142/S0218301316300010",
    journal = "Int. J. Mod. Phys. E",
    volume = "25",
    pages = "1630001",
    year = "2016"
}

@article{Swanson:2003tb,
    author = "Swanson, Eric S.",
    title = "{Short range structure in the X(3872)}",
    eprint = "hep-ph/0311229",
    archivePrefix = "arXiv",
    reportNumber = "JLAB-THY-03-227",
    doi = "10.1016/j.physletb.2004.03.033",
    journal = "Phys. Lett. B",
    volume = "588",
    pages = "189--195",
    year = "2004"
}

@article{Lebed:2016hpi,
    author = "Lebed, Richard F. and Mitchell, Ryan E. and Swanson, Eric S.",
    title = "{Heavy-Quark QCD Exotica}",
    eprint = "1610.04528",
    archivePrefix = "arXiv",
    primaryClass = "hep-ph",
    doi = "10.1016/j.ppnp.2016.11.003",
    journal = "Prog. Part. Nucl. Phys.",
    volume = "93",
    pages = "143--194",
    year = "2017"
}

@article{Ali:2017jda,
    author = {Ali, Ahmed and Lange, Jens S{\"o}ren and Stone, Sheldon},
    title = "{Exotics: Heavy Pentaquarks and Tetraquarks}",
    eprint = "1706.00610",
    archivePrefix = "arXiv",
    primaryClass = "hep-ph",
    reportNumber = "DESY-17-071",
    doi = "10.1016/j.ppnp.2017.08.003",
    journal = "Prog. Part. Nucl. Phys.",
    volume = "97",
    pages = "123--198",
    year = "2017"
}

@article{Guo:2017jvc,
    author = "Guo, Feng-Kun and Hanhart, Christoph and Mei{\ss}ner, Ulf-G. and Wang, Qian and Zhao, Qiang and Zou, Bing-Song",
    title = "{Hadronic molecules}",
    eprint = "1705.00141",
    archivePrefix = "arXiv",
    primaryClass = "hep-ph",
    doi = "10.1103/RevModPhys.90.015004",
    journal = "Rev. Mod. Phys.",
    volume = "90",
    number = "1",
    pages = "015004",
    year = "2018",
    note = "[Erratum: Rev.Mod.Phys. 94, 029901 (2022)]"
}

@article{Dong:2021bvy,
    author = "Dong, Xiang-Kun and Guo, Feng-Kun and Zou, Bing-Song",
    title = "{A survey of heavy{\textendash}heavy hadronic molecules}",
    eprint = "2108.02673",
    archivePrefix = "arXiv",
    primaryClass = "hep-ph",
    doi = "10.1088/1572-9494/ac27a2",
    journal = "Commun. Theor. Phys.",
    volume = "73",
    number = "12",
    pages = "125201",
    year = "2021"
}

@article{Ishii:2006ec,
    author = "Ishii, N. and Aoki, S. and Hatsuda, T.",
    title = "{The Nuclear Force from Lattice QCD}",
    eprint = "nucl-th/0611096",
    archivePrefix = "arXiv",
    reportNumber = "UTCCS-P-30, TKYNT-06-19, UTHEP-534",
    doi = "10.1103/PhysRevLett.99.022001",
    journal = "Phys. Rev. Lett.",
    volume = "99",
    pages = "022001",
    year = "2007"
}

@article{Guo:2019twa,
    author = "Guo, Feng-Kun and Liu, Xiao-Hai and Sakai, Shuntaro",
    title = "{Threshold cusps and triangle singularities in hadronic reactions}",
    eprint = "1912.07030",
    archivePrefix = "arXiv",
    primaryClass = "hep-ph",
    doi = "10.1016/j.ppnp.2020.103757",
    journal = "Prog. Part. Nucl. Phys.",
    volume = "112",
    pages = "103757",
    year = "2020"
}

@article{Machleidt:1987hj,
    author = "Machleidt, R. and Holinde, K. and Elster, C.",
    title = "{The Bonn Meson Exchange Model for the Nucleon Nucleon Interaction}",
    doi = "10.1016/S0370-1573(87)80002-9",
    journal = "Phys. Rept.",
    volume = "149",
    pages = "1--89",
    year = "1987"
}

@article{Liu:2018zzu,
    author = "Liu, Ming-Zhu and Peng, Fang-Zheng and S{\'a}nchez S{\'a}nchez, Mario and Valderrama, Manuel Pavon",
    title = "{Heavy-quark symmetry partners of the $P_c(4450)$ pentaquark}",
    eprint = "1811.03992",
    archivePrefix = "arXiv",
    primaryClass = "hep-ph",
    doi = "10.1103/PhysRevD.98.114030",
    journal = "Phys. Rev. D",
    volume = "98",
    number = "11",
    pages = "114030",
    year = "2018"
}

@article{Guo:2013sya,
    author = "Guo, Feng-Kun and Hidalgo-Duque, Carlos and Nieves, Juan and Valderrama, Manuel Pavon",
    title = "{Consequences of Heavy Quark Symmetries for Hadronic Molecules}",
    eprint = "1303.6608",
    archivePrefix = "arXiv",
    primaryClass = "hep-ph",
    doi = "10.1103/PhysRevD.88.054007",
    journal = "Phys. Rev. D",
    volume = "88",
    pages = "054007",
    year = "2013"
}

@article{Liu:2019stu,
    author = "Liu, Ming-Zhu and Wu, Tian-Wei and Pavon Valderrama, Manuel and Xie, Ju-Jun and Geng, Li-Sheng",
    title = "{Heavy-quark spin and flavor symmetry partners of the X(3872) revisited: What can we learn from the one boson exchange model?}",
    eprint = "1902.03044",
    archivePrefix = "arXiv",
    primaryClass = "hep-ph",
    doi = "10.1103/PhysRevD.99.094018",
    journal = "Phys. Rev. D",
    volume = "99",
    number = "9",
    pages = "094018",
    year = "2019"
}

@article{Liu:2019tjn,
    author = "Liu, Ming-Zhu and Pan, Ya-Wen and Peng, Fang-Zheng and S{\'a}nchez S{\'a}nchez, Mario and Geng, Li-Sheng and Hosaka, Atsushi and Pavon Valderrama, Manuel",
    title = "{Emergence of a complete heavy-quark spin symmetry multiplet: seven molecular pentaquarks in light of the latest LHCb analysis}",
    eprint = "1903.11560",
    archivePrefix = "arXiv",
    primaryClass = "hep-ph",
    doi = "10.1103/PhysRevLett.122.242001",
    journal = "Phys. Rev. Lett.",
    volume = "122",
    number = "24",
    pages = "242001",
    year = "2019"
}

@article{Chen:2015loa,
    author = "Chen, Rui and Liu, Xiang and Li, Xue-Qian and Zhu, Shi-Lin",
    title = "{Identifying exotic hidden-charm pentaquarks}",
    eprint = "1507.03704",
    archivePrefix = "arXiv",
    primaryClass = "hep-ph",
    doi = "10.1103/PhysRevLett.115.132002",
    journal = "Phys. Rev. Lett.",
    volume = "115",
    number = "13",
    pages = "132002",
    year = "2015"
}

@article{Tornqvist:1993ng,
    author = "Tornqvist, Nils A.",
    title = "{From the deuteron to deusons, an analysis of deuteron - like meson meson bound states}",
    eprint = "hep-ph/9310247",
    archivePrefix = "arXiv",
    reportNumber = "HU-SEFT-R-1993-12",
    doi = "10.1007/BF01413192",
    journal = "Z. Phys. C",
    volume = "61",
    pages = "525--537",
    year = "1994"
}

@article{Meng:2022ozq,
    author = "Meng, Lu and Wang, Bo and Wang, Guang-Juan and Zhu, Shi-Lin",
    title = "{Chiral perturbation theory for heavy hadrons and chiral effective field theory for heavy hadronic molecules}",
    eprint = "2204.08716",
    archivePrefix = "arXiv",
    primaryClass = "hep-ph",
    doi = "10.1016/j.physrep.2023.04.003",
    journal = "Phys. Rept.",
    volume = "1019",
    pages = "1--149",
    year = "2023"
}

@article{Lednicky:1981su,
    author = "Lednicky, R. and Lyuboshits, V. L.",
    title = "{Final State Interaction Effect on Pairing Correlations Between Particles with Small Relative Momenta}",
    reportNumber = "JINR-E2-81-453",
    journal = "Yad. Fiz.",
    volume = "35",
    pages = "1316--1330",
    year = "1981"
}

@article{Briceno:2017max,
    author = "Briceno, Raul A. and Dudek, Jozef J. and Young, Ross D.",
    title = "{Scattering processes and resonances from lattice QCD}",
    eprint = "1706.06223",
    archivePrefix = "arXiv",
    primaryClass = "hep-lat",
    reportNumber = "JLAB-THY-17-2495, ADP-17-28-T1034",
    doi = "10.1103/RevModPhys.90.025001",
    journal = "Rev. Mod. Phys.",
    volume = "90",
    number = "2",
    pages = "025001",
    year = "2018"
}

@article{Dong:2017gaw,
    author = "Dong, Yubing and Faessler, Amand and Lyubovitskij, Valery E.",
    title = "{Description of heavy exotic resonances as molecular states using phenomenological Lagrangians}",
    doi = "10.1016/j.ppnp.2017.01.002",
    journal = "Prog. Part. Nucl. Phys.",
    volume = "94",
    pages = "282--310",
    year = "2017"
}

@article{Esposito:2016noz,
    author = "Esposito, A. and Pilloni, A. and Polosa, A. D.",
    title = "{Multiquark Resonances}",
    eprint = "1611.07920",
    archivePrefix = "arXiv",
    primaryClass = "hep-ph",
    reportNumber = "JLAB-THY-16-2301",
    doi = "10.1016/j.physrep.2016.11.002",
    journal = "Phys. Rept.",
    volume = "668",
    pages = "1--97",
    year = "2017"
}

@article{Karliner:2017qhf,
    author = "Karliner, Marek and Rosner, Jonathan L. and Skwarnicki, Tomasz",
    title = "{Multiquark States}",
    eprint = "1711.10626",
    archivePrefix = "arXiv",
    primaryClass = "hep-ph",
    doi = "10.1146/annurev-nucl-101917-020902",
    journal = "Ann. Rev. Nucl. Part. Sci.",
    volume = "68",
    pages = "17--44",
    year = "2018"
}

@article{Li:2012ss,
    author = "Li, Ning and Sun, Zhi-Feng and Liu, Xiang and Zhu, Shi-Lin",
    title = "{Coupled-channel analysis of the possible $D^{(*)}D^{(*)}, \overline{B}^{(*)}\overline{B}^{(*)}$ and $D^{(*)}\overline{B}^{(*)}$ molecular states}",
    eprint = "1211.5007",
    archivePrefix = "arXiv",
    primaryClass = "hep-ph",
    doi = "10.1103/PhysRevD.88.114008",
    journal = "Phys. Rev. D",
    volume = "88",
    number = "11",
    pages = "114008",
    year = "2013"
}

@article{Yalikun:2021bfm,
    author = "Yalikun, Nijiati and Lin, Yong-Hui and Guo, Feng-Kun and Kamiya, Yuki and Zou, Bing-Song",
    title = "{Coupled-channel effects of the {\ensuremath{\Sigma}}c(*)D{\textasciimacron}(*)-{\ensuremath{\Lambda}}c(2595)D{\textasciimacron} system and molecular nature of the Pc pentaquark states from one-boson exchange model}",
    eprint = "2109.03504",
    archivePrefix = "arXiv",
    primaryClass = "hep-ph",
    doi = "10.1103/PhysRevD.104.094039",
    journal = "Phys. Rev. D",
    volume = "104",
    number = "9",
    pages = "094039",
    year = "2021"
}

@article{Yang:2020nrt,
    author = "Yang, Zhi and Cao, Xu and Guo, Feng-Kun and Nieves, Juan and Valderrama, Manuel Pavon",
    title = "{Strange molecular partners of the $Z_c$(3900) and $Z_c$(4020)}",
    eprint = "2011.08725",
    archivePrefix = "arXiv",
    primaryClass = "hep-ph",
    doi = "10.1103/PhysRevD.103.074029",
    journal = "Phys. Rev. D",
    volume = "103",
    number = "7",
    pages = "074029",
    year = "2021"
}

@article{Epelbaum:2008ga,
    author = "Epelbaum, Evgeny and Hammer, Hans-Werner and Meissner, Ulf-G.",
    title = "{Modern Theory of Nuclear Forces}",
    eprint = "0811.1338",
    archivePrefix = "arXiv",
    primaryClass = "nucl-th",
    reportNumber = "HISKP-TH-08-18, FZJ-IKP-TH-2008-20",
    doi = "10.1103/RevModPhys.81.1773",
    journal = "Rev. Mod. Phys.",
    volume = "81",
    pages = "1773--1825",
    year = "2009"
}

@article{Liu:2023wfo,
    author = "Liu, Zhi-Wei and Lu, Jun-Xu and Liu, Ming-Zhu and Geng, Li-Sheng",
    title = "{Distinguishing the spins of Pc(4440) and Pc(4457) with femtoscopic correlation functions}",
    eprint = "2305.19048",
    archivePrefix = "arXiv",
    primaryClass = "hep-ph",
    doi = "10.1103/PhysRevD.108.L031503",
    journal = "Phys. Rev. D",
    volume = "108",
    number = "3",
    pages = "L031503",
    year = "2023"
}

@article{ALICE:2026bgl,
    author = "Abualrob, Ibrahim Jaser and others",
    collaboration = "ALICE",
    title = "{One- and three-dimensional identical charged-kaon femtoscopic correlations in Pb-Pb collisions at sNN=5.02TeV}",
    eprint = "2601.15054",
    archivePrefix = "arXiv",
    primaryClass = "nucl-ex",
    reportNumber = "CERN-EP-2025-236",
    doi = "10.1103/z6kw-l1pc",
    journal = "Phys. Rev. C",
    volume = "114",
    number = "3",
    pages = "034005",
    year = "2026"
}

@article{Wang:2025sic,
    author = "Wang, Zhi-Gang",
    title = "{Review of the QCD sum rules for exotic states}",
    eprint = "2502.11351",
    archivePrefix = "arXiv",
    primaryClass = "hep-ph",
    doi = "10.15302/frontphys.2026.016300",
    journal = "Front. Phys. (Beijing)",
    volume = "21",
    number = "1",
    pages = "016300",
    year = "2026"
}

@article{Doring:2025sgb,
    author = {D{\"o}ring, Michael and Haidenbauer, Johann and Mai, Maxim and Sato, Toru},
    title = "{Dynamical coupled-channel models for hadron dynamics}",
    eprint = "2505.02745",
    archivePrefix = "arXiv",
    primaryClass = "nucl-th",
    reportNumber = "JLAB-THY-25-4298",
    doi = "10.1016/j.ppnp.2025.104213",
    journal = "Prog. Part. Nucl. Phys.",
    volume = "146",
    pages = "104213",
    year = "2026"
}

@article{Lu:2021gsb,
    author = "Lu, Jun-Xu and Wang, Chun-Xuan and Xiao, Yang and Geng, Li-Sheng and Meng, Jie and Ring, Peter",
    title = "{Accurate Relativistic Chiral Nucleon-Nucleon Interaction up to Next-to-Next-to-Leading Order}",
    eprint = "2111.07766",
    archivePrefix = "arXiv",
    primaryClass = "nucl-th",
    doi = "10.1103/PhysRevLett.128.142002",
    journal = "Phys. Rev. Lett.",
    volume = "128",
    number = "14",
    pages = "142002",
    year = "2022"
}

@article{Yuan:2025pnt,
    author = "Yuan, Qi-Wei and Wu, Qi and Liu, Ming-Zhu",
    title = "{B-meson decays to vector charmoniumlike states and a K meson: The role of final-state interactions}",
    eprint = "2504.11121",
    archivePrefix = "arXiv",
    primaryClass = "hep-ph",
    doi = "10.1103/xrjr-zxpz",
    journal = "Phys. Rev. D",
    volume = "111",
    number = "11",
    pages = "114035",
    year = "2025"
}

@article{Jia:2024pyb,
    author = "Jia, Cai-Ping and Jiang, Hua-Yu and Wang, Jian-Peng and Yu, Fu-Sheng",
    title = "{Final-state rescattering mechanism of charmed baryon decays}",
    eprint = "2408.14959",
    archivePrefix = "arXiv",
    primaryClass = "hep-ph",
    doi = "10.1007/JHEP11(2024)072",
    journal = "JHEP",
    volume = "11",
    pages = "072",
    year = "2024"
}

@article{Duan:2024zjv,
    author = {Duan, Zhu-Ding and Wang, Jian-Peng and Li, Run-Hui and L{\"u}, Cai-Dian and Yu, Fu-Sheng},
    title = "{Final-state rescattering mechanism in bottom-baryon decays}",
    eprint = "2412.20458",
    archivePrefix = "arXiv",
    primaryClass = "hep-ph",
    doi = "10.1007/JHEP09(2025)160",
    journal = "JHEP",
    volume = "09",
    pages = "160",
    year = "2025"
}

@article{Wang:2012mf,
    author = "Wang, Qian and Li, Gang and Zhao, Qiang",
    title = "{Open charm effects in the explanation of the long-standing '$\rho\pi$ puzzle'}",
    eprint = "1201.1681",
    archivePrefix = "arXiv",
    primaryClass = "hep-ph",
    doi = "10.1103/PhysRevD.85.074015",
    journal = "Phys. Rev. D",
    volume = "85",
    pages = "074015",
    year = "2012"
}

@article{Zhao:2010mm,
    author = "Zhao, Qiang",
    title = "{Understanding the radiative decays of vector charmonia to light pseudoscalar mesons}",
    eprint = "1012.1165",
    archivePrefix = "arXiv",
    primaryClass = "hep-ph",
    doi = "10.1016/j.physletb.2011.01.043",
    journal = "Phys. Lett. B",
    volume = "697",
    pages = "52--57",
    year = "2011"
}

\end{document}